\documentclass[twocolumn]{aastex63}

\usepackage{multirow}
\usepackage{pbox}

\newcommand{\esc}{erg cm$^{-2}$ s$^{-1}$}

\newcommand{\kms}{km s$^{-1}$}

\newcommand{\mstar}{$M_{\star}$}

\newcommand{\mdust}{$M_{\rm dust}$}
\newcommand{\kkmspc}{K~km~s$^{-1}$~pc$^2$}

\newcommand{\jykms}{Jy~km~s$^{-1}$}
\newcommand{\ci}{[C\,{\footnotesize I}]}

\newcommand{\co}{CO}
\newcommand{\cione}{[C\,{\footnotesize I}]$(^3P_1\,-\,^{3}P_0)$}
\newcommand{\citwo}{[C\,{\footnotesize I}]$(^3P_2\,-\, ^{3}P_1)$}
\newcommand{\ciplus}{[C\,{\footnotesize II}]}
\newcommand{\cofive}{CO\,$(5-4)$}
\newcommand{\coseven}{CO\,$(7-6)$}
\newcommand{\cofour}{CO\,$(4-3)$}
\newcommand{\cothree}{CO\,$(3-2)$}
\newcommand{\cotwo}{CO\,$(2-1)$}
\newcommand{\coone}{CO\,$(1-0)$}
\newcommand{\lprimeci}{$L'_{\mathrm{[C\,\scriptscriptstyle{I}\scriptstyle{]}}^3P_1\,-\, ^3P_0}$}
\newcommand{\lci}{$L_{\mathrm{[C\,\scriptscriptstyle{I}\scriptstyle{]}}^3P_1\,-\, ^3P_0}$}

\newcommand{\lcione}{$L_{\mathrm{[C\,\scriptscriptstyle{I}\scriptstyle{]}}^3P_1\,-\, ^3P_0}$}
\newcommand{\lcitwo}{$L_{\mathrm{[C\,\scriptscriptstyle{I}\scriptstyle{]}}^3P_2\,-\, ^3P_1}$}

\newcommand{\lprimecione}{$L'_{\mathrm{[C\,\scriptscriptstyle{I}\scriptstyle{]}}^3P_1\,-\, ^3P_0}$}

\newcommand{\lprimecitwo}{$L'_{\mathrm{[C\,\scriptscriptstyle{I}\scriptstyle{]}}^3P_2\,-\, ^3P_1}$}
\newcommand{\lprimecotwo}{$L'_{\rm CO(2-1)}$}
\newcommand{\lprimecoone}{$L'_{\rm CO(1-0)}$}
\newcommand{\lprimecothree}{$L'_{\rm CO(3-2)}$}
\newcommand{\lprimecofour}{$L'_{\rm CO(4-3)}$}
\newcommand{\lprimecofive}{$L'_{\rm CO(5-4)}$}
\newcommand{\lprimecoseven}{$L'_{\rm CO(7-6)}$}

\newcommand{\lcothree}{$L_{\rm CO(3-2)}$}
\newcommand{\lcofour}{$L_{\rm CO(4-3)}$}
\newcommand{\lcofive}{$L_{\rm CO(5-4)}$}
\newcommand{\lcoseven}{$L_{\rm CO(7-6)}$}
\newcommand{\ici}{$I_{\mathrm{[C\,\scriptscriptstyle{I}\scriptstyle{]}}^3P_1\,-\, ^3P_0}$}
\newcommand{\icitwo}{$I_{\mathrm{[C\,\scriptscriptstyle{I}\scriptstyle{]}}^3P_2\,-\, ^3P_1}$}
\newcommand{\lir}{$L_{\rm IR}$}
\newcommand{\lfir}{$L_{\rm FIR}$}
\newcommand{\lsun}{$L_{\odot}$}

\received{2019}
\revised{2019}
\submitjournal{ApJ}

\shorttitle{The properties of the interstellar medium as traced by the
  neutral atomic carbon \textsc{[C\,{\footnotesize I}]}}
\shortauthors{Valentino et al.}
\begin{document}

\title{The properties of the interstellar medium of galaxies across time as traced by the
  neutral atomic carbon \textsc{[C\,{\footnotesize I}]}}

%% Authors

\correspondingauthor{Francesco Valentino}
\email{francesco.valentino@nbi.ku.dk}

\author[0000-0001-6477-4011]{Francesco Valentino}
\affil{Cosmic Dawn Center (DAWN)}
\affil{Niels Bohr Institute, University of
  Copenhagen, Lyngbyvej 2, DK-2100 Copenhagen \O}

\author[0000-0002-4872-2294]{Georgios E. Magdis}
\affil{Cosmic Dawn Center (DAWN)}
\affil{Niels Bohr Institute, University of
 Copenhagen, Lyngbyvej 2, DK-2100 Copenhagen \O}
\affil{DTU-Space,
  Technical University of Denmark, Elektrovej 327, DK-2800 Kgs.\
  Lyngby}
\affil{Institute for Astronomy, Astrophysics, Space Applications and
Remote Sensing, National Observatory of Athens, GR-15236 Athens, Greece}

\author[0000-0002-3331-9590]{Emanuele Daddi}
\affil{Laboratoire AIM-Paris-Saclay, CEA/DSM-CNRS-Universit\'{e} Paris
Diderot, Irfu/Service d'Astrophysique, CEA Saclay, Orme des Merisiers, F-91191
Gif sur Yvette, France}

\author[0000-0001-9773-7479]{Daizhong Liu}
\affil{Max Planck Institute for Astronomy, K\"{o}nigstuhl 17, D-69117 Heidelberg, Germany}

\author[0000-0002-6290-3198]{Manuel Aravena}
\affil{N\'{u}cleo de Astronom\'{i}a, Facultad de Ingenier\'{i}a y
Ciencias, Universidad Diego Portales, Av. Ejército 441, Santiago, Chile}

\author[0000-0002-5743-0250]{Fr\'{e}d\'{e}ric Bournaud}
\affil{Laboratoire AIM-Paris-Saclay, CEA/DSM-CNRS-Universit\'{e} Paris
 Diderot, Irfu/Service d'Astrophysique, CEA Saclay, Orme des Merisiers, F-91191
 Gif sur Yvette, France}

\author[0000-0001-9197-7623]{Isabella Cortzen}
\affil{Cosmic Dawn Center (DAWN)}
\affil{Niels Bohr Institute, University of 
  Copenhagen, Lyngbyvej 2, DK-2100 Copenhagen \O}

\author[0000-0003-0007-2197]{Yu Gao}
\affil{Purple Mountain Observatory \& Key Laboratory for Radio Astronomy, Chinese Academy of Sciences, 10 Yuanhua Road, 
Nanjing 210033, People's Republic of China}

\author[0000-0002-8412-7951]{Shuowen Jin}
\affil{Instituto de Astrof\'{i}sica de Canarias (IAC), E-38205 La
  Laguna, Tenerife, Spain}
\affil{Universidad de La Laguna, Dpto. Astrof\'{i}sica, E-38206 La Laguna, Tenerife, Spain}

\author[0000-0001-5414-5131]{St\'{e}phanie Juneau}
\affil{National Optical Astronomy Observatory, 950 North Cherry
  Avenue, Tucson, AZ 85719, USA}

\author[0000-0001-9187-3605]{Jeyhan S. Kartaltepe}
\affil{School of Physics and Astronomy, Rochester Institute of
 Technology, 84 Lomb Memorial Drive, Rochester NY 14623, USA}

\author[0000-0002-5588-9156]{Vasily Kokorev}
\affil{Cosmic Dawn Center (DAWN)}
\affil{Niels Bohr Institute, University of
  Copenhagen, Lyngbyvej 2, DK-2100 Copenhagen \O}

\author[0000-0002-9888-0784]{Min-Young Lee}
\affil{Korea Astronomy and Space Science Institute, 776 Daedeokdae-ro, 34055 Daejeon, Republic of Korea}
\affil{Max-Planck-Institut f\"ur Radioastronomie, Auf dem H\"ugel 69, 53121 Bonn, Germany}

\author[0000-0003-3229-2899]{Suzanne C. Madden}
\affil{Laboratoire AIM-Paris-Saclay, CEA/DSM-CNRS-Universit\'{e} Paris
 Diderot, Irfu/Service d'Astrophysique, CEA Saclay, Orme des Merisiers, F-91191
 Gif sur Yvette, France}

\author[0000-0002-7064-4309]{Desika Narayanan}
\affil{Cosmic Dawn Center (DAWN)}
\affil{Department of Astronomy, University of Florida, 211 Bryant Space Sciences Center, Gainesville, FL 32611 USA}
\affil{University of Florida Informatics Institute, 432 Newell Drive, CISE Bldg E251, Gainesville, FL 32611}

\author[0000-0003-1151-4659]{Gerg\"{o} Popping}
\affil{European Southern Observatory, Karl-Schwarzschild-Strasse 2, D-85745, Garching, Germany}

\author[0000-0001-9369-1805]{Annagrazia Puglisi}
\affil{Laboratoire AIM-Paris-Saclay, CEA/DSM-CNRS-Universit\'{e} Paris
 Diderot, Irfu/Service d'Astrophysique, CEA Saclay, Orme des Merisiers, F-91191
 Gif sur Yvette, France}

%% Abstract
\begin{abstract}
We report ALMA
observations of the neutral atomic carbon transitions \ci\ and
multiple CO lines in a sample of $\sim30$ main sequence
galaxies at $z\sim1$, including novel information on \citwo\ and
\coseven\ for $7$ of such normal objects. We complement our observations with a
collection of $>200$ galaxies with coverage of similar transitions,
spanning the $z=0-4$ redshift interval and a variety of ambient
conditions from local to high-redshift starbursts. We
find systematic variations in the \ci/IR and \ci/high-$J_{\rm upper}$ ($J_{\rm upper}=7$) CO
luminosity ratios among the various samples. We interpret these differences as
increased dense molecular gas fractions and star formation
efficiencies in the strongest high-redshift starbursts with respect to 
normal main sequence galaxies. We further report constant
\lprimecitwo/\lprimecione\ ratios across the galaxy populations and
redshifts, suggesting that gas temperatures $T_{\rm exc}$ traced by
\ci\ do not strongly vary. We find only a mild
  correlation with $T_{\rm dust}$ and that, generally,
$T_{\rm exc} \lesssim T_{\rm dust}$. We fit
the line ratios with classical photodissociation
region models, retrieving consistently larger densities and
intensities of the UV radiation fields in sub-mm galaxies than in main
sequence and local objects. However, these simple models fall short in
representing the complexity of a multi-phase interstellar
medium and should be treated with caution. Finally, we compare our
observations with the Santa Cruz
semi-analytical model of galaxy evolution, recently extended to
simulate sub-mm emission. While we confirm the
success in reproducing the \co\ lines, we find systematically
larger \ci\ luminosities at fixed IR luminosity than predicted theoretically. This 
highlights
the necessity of improving our understanding of the mechanisms
regulating the \ci\ emission on galactic scales. We release
our data compilation to the community.

\end{abstract}

%% Keywords 

\section{Introduction}
\label{sec:introduction}
An accurate description of the physical mechanisms regulating gas in
galaxies is paramount to reach a complete understanding of how these
systems evolve with cosmic time. The recent advent of powerful
interferometers such as the NOrthern Extended Millimeter Array (NOEMA)
and, especially, the Atacama Large Millimeter Array (ALMA) have played
a major role in this regard, opening a window not only on traditional
molecular gas tracers, such as $^{12}$CO and dust
\citep[e.g.,][]{magdis_2012b, carilli_2013,
  bolatto_2013, scoville_2014} in large samples of distant galaxies, but also
on lines previously inaccessible because of their intrinsic
faintness. These facilities have also allowed the study of galaxies at
unprecedented redshifts and
those representative of the bulk of the population, in addition to the
extremely bright starbursts and
sub-mm galaxies (SMGs). Particular attention has been given to
alternative proxies for the total molecular gas mass in galaxies,
highly desirable to complement \co\ and dust and to break the known
degeneracies hampering these tracers \citep{zanella_2018,
  cortzen_2019}. However, multiple elements and molecules have now been
detected in distant systems, allowing us to study their ionization conditions,
chemistry, metallicity, densities and temperatures.\\

In a previous work \citep[][V18 hereafter]{valentino_2018}, we
presented results on the lowest neutral carbon transition \cione\
($\nu_{\rm rest}=492.161$~ GHz) in normal galaxies at $z\sim1.2$ that lie on
the so called ``main sequence'' in the
stellar mass - star formation rate plane
(\mstar-SFR, \citealt{noeske_2007, elbaz_2007, daddi_2007, magdis_2010}; see also
\citealt{bourne_2019, brisbin_2019, lamarche_2019}). The use of
\ci\ has theoretical and observational roots that have been
deepened over the years, revealing potential advantages over
alternative tracers \citep[e.g.,][Madden et al. in
prep.]{papadopoulos_2004}. In V18 we showed that \cione\ and low-$J$
\co\ line emissions correlate on global scales irrespectively of the
redshift and galaxy type, bridging previous observations of local
infrared (IR) luminous objects \citep[e.g.,][]{gerin_2000, papadopoulos_2004,
  liu_2015, israel_2015, kamenetzky_2016, jiao_2017, jiao_2019} and
high-redshift SMGs \citep[][to mention some recent
efforts]{walter_2011, alaghband-zadeh_2013, yang_2017, bothwell_2017,
  andreani_2018, canameras_2018, nesvadba_2018}. Moreover, we reported
systematic variations of the \lprimecione/\lir\ luminosity ratio of normal
main sequence galaxies from the starbursting SMG population,
consistently with results based on \co\ \citep[e.g.,][]{daddi_2010b,
  magdis_2012, genzel_2015, tacconi_2018}, supporting the existence of
different star formation regimes characterized by varying star
formation efficiencies ($\mathrm{SFE = SFR}/M_\star$). This also
resulted in different \ci\ abundances in main sequence, starburst
and SMG galaxies, naturally following the standard assumptions on the
$\alpha_{\rm CO}$ and dust-to-gas conversion factors.\\

Here we move from the ground we laid in our previous work (i) expanding
our analysis to larger samples that became available during the last year,
and (ii) simultaneously studying multiple line transitions, allowing
us to study the properties of the interstellar medium across redshift
and galaxy types. In particular, we introduce new ALMA observations of the
excited \citwo\ line ($\nu_{\rm rest} =809.344$~GHz) in $30$\% of the
main sequence galaxies that we presented in V18, giving simultaneous
access to the \coseven\ transition. These observations open a view both on the
excited \ci\ gas, allowing for a direct estimate of the gas temperature
via the \citwo/\cione\ ratio \citep{stutzki_1997, schneider_2003, weiss_2003,
  papadopoulos_2004}, and on the dense and warm gas phases so far
explored only in the brightest galaxies at high redshift
\citep[e.g.,][]{yang_2017, canameras_2018, apostolovski_2019}. The availability of multiple
line ratios for sizable and controlled samples of high-redshift
galaxies is also the basis for modeling the \ci, \co, and dust
emission, a historically complicated endeavor especially for what
concerns \ci. Classical monodimensional photodissociation region models predict
the \ci\ emission to arise only from a thin layer in between \ciplus\ and
\co\ \citep{tielens_1985, kaufman_1999}, struggling to reproduce fully
concomitant \ci\ and \co\ emission in local giant molecular clouds
\citep{keene_1996, ojha_2001, ikeda_2002}. More recent refinement,
including non-equilibrium chemistry \citep{storzer_1997}, turbulent
mixing \citep{xie_1995, glover_2015}, clumpy geometries \citep{stutzki_1998}, and
the effect of cosmic rays \citep{papadopoulos_2004, papadopoulos_2018,
  bisbas_2015, bisbas_2017} and 3D geometry \citep{bisbas_2012} have
been more successful in this sense. The addition of extra
heating mechanisms \citep[e.g., shocks,][]{lee_2019}, or
radically different approaches \citep[e.g., large velocity gradients,
LVG,][]{young_1991} have been recently successful in reproducing the
interstellar medium (ISM) conditions in local resolved star forming regions or \co+\ci\
emission in nuclear starbursts \citep{israel_2015}, detecting
multiple phases traced by different line transitions. However, such
modeling requires large amount of data tracing the various ISM
components, which become progressively hard to collect at increasing redshifts
or for faint galaxies. A tradeoff between model complexity and its
applicability is what we aim at in this work.\\  

Finally, the availability of a large compilation of galaxies with \ci\
detections allows one to insert this emission line
in the cosmological context of galaxy evolution. Recent works have been
focusing on the modeling of the \co\ and \ciplus\ emission, given
their intrinsic brightness and coverage up to extremely high redshift
\citep[see][and references therein for a recent review]{olsen_2018},
reaping the first rewards of such effort. Less attention has been dedicated
to \ci, but models are quickly filling the gap. Here we focus on a
recent implementation of the sub-mm lines emission modeling onto the
Santa Cruz semi-analytical model described in \citet{popping_2019}. We
will show how the fiducial model compares with the observed \ci,
\co, and IR luminosities and how these observables can be
inserted in the empirical frame of known scaling relations across
redshifts.\\

This paper is organized as follows. In Section \ref{sec:data}, we
present the new ALMA data targeting \citwo+\coseven\ in main sequence
galaxies at $z\sim1.2$, along with the description of the data
compilation we assembled from the literature. Section \ref{sec:results} includes the main
observational results of our work and their interpretation in the frame of the
empirical scaling relations. In Section \ref{sec:discussion}, we apply
a simple photodissociation region model to interpret the
observed trends, we discuss its limitations, and we introduce the
comparison with the semi-analytical modeling mentioned above. Section
\ref{sec:conclusions} summarizes the results of our work. The whole
data compilation we assembled for this work is made publicly available in an
electronic format in the online version or by requesting it to the
contact author. Unless stated otherwise, we assume a $\Lambda$CDM cosmology with
$\Omega_{\rm m} = 0.3$, $\Omega_{\rm \Lambda} = 0.7$, and $H_0 = 70$
km s$^{-1}$ Mpc$^{-1}$ and a Chabrier initial mass function
\citep[IMF,][]{chabrier_2003}. All magnitudes are expressed in the AB system.  
All the literature data have been homogenized with our conventions.

\section{Sample and observations}
\label{sec:data}

The galaxies we study here largely overlap with the sample of main
sequence galaxies presented in \cite{valentino_2018}. In this work, we
also present new observations of the \citwo\ and \coseven\ transitions for 
7 objects from V18 observed during ALMA Cycle 6. We further add the recent
observations of main sequence galaxies by \cite{bourne_2019} and
\cite{popping_2017}, and the
sub-mm galaxies listed in \cite{yang_2017, andreani_2018,
  canameras_2018, nesvadba_2018, dannerbauer_2019, jin_2019}. In total we
retrieve information about \ci\ emission in 217 galaxies\footnote{Our
  collection refers to published material prior to May 2019, to the
  best of our knowledge. Other works have been
    brought to our attention after we conducted our analysis
    \citep[e.g.,][]{gullberg_2016, strandet_2017, lelli_2018,
      man_2019}. These and further results will be included in future
    versions of this database.}. A certain degree of
  inhomogeneity is inherent in the nature of such a large literature
  compilation (e.g., different selections, single dish vs interferometric
  observations, flux extraction, modeling). Whenever possible, we homogenized the measurements
  adopting a single approach, for example when fitting the far-IR SED
 (see below, Appendix \ref{sec:app:sed_literature}, and the supplementary material).
Here we briefly summarize the salient
properties of each sample, referring the reader to
\cite{valentino_2018} and the original papers for further
details. Statistics of the data collection are reported in Table
\ref{tab:compilation}.
  
 \begin{deluxetable*}{lccccc}
   \tabletypesize{\small}
   \tablecolumns{6}
   \tablecaption{Statistics of the data compilation\label{tab:compilation}.}
   \smallskip
   \tablehead{
     \colhead{Reference} & 
     \colhead{\cione}&
     \colhead{\citwo}&
     \colhead{$J_{\rm upper}< 3$}&
     \colhead{$3\leq J_{\rm upper}\leq 6$} & 
     \colhead{$J_{\rm upper}\geq 7$}
   }
   \startdata
   \rule{0pt}{-10ex}& & & & &\\
   \multicolumn{6}{c}{High-redshift main sequence galaxies}\smallskip\\
   \hline
   This work, \cite{valentino_2018}& 20 (3)& 7 (0) & 11 (0) & 19 (0)& 7 (0)\smallskip\\
   \cite{bourne_2019}& 6 (4)& \nodata& \nodata& 4 (5)& \nodata\smallskip\\
   \cite{popping_2017,talia_2018}& 1 (0)& \nodata& \nodata& 1 (0)& \nodata\smallskip\\
   \hline
   Total& 27 (7)& 7 (0)& 11 (0)& 24 (5)& 7 (0)\\ 
   \rule{0pt}{-10ex}& & & & &\\
   \multicolumn{6}{c}{Local IR-luminous galaxies}\smallskip\\
   \hline
   \cite{liu_2015, kamenetzky_2016}$^\dagger$& 32 (114)& 126 (20)& 29 (0)& 56 (90)& 104 (42)\smallskip\\
   \hline
   \rule{0pt}{-10ex}& & & & &\\
   \multicolumn{6}{c}{High-redshift SMGs and QSOs}\smallskip\\
   \hline
   \cite{walter_2011, alaghband-zadeh_2013};& \multirow{2}{*}{17 (4)}& \multirow{2}{*}{11 (7)}& \multirow{2}{*}{\nodata}& \multirow{2}{*}{22 (0)}& \multirow{2}{*}{12 (1)}\\
   Cortzen et al. (submitted)& & & & & \smallskip\\
   \cite{bothwell_2017}& 9 (4)& \nodata& 9 (0)& 13 (0)& \nodata\smallskip\\
   \cite{canameras_2018, nesvadba_2018};& \multirow{2}{*}{7 (0)}& \multirow{2}{*}{8 (0)}& \multirow{2}{*}{5 (0)}&
   \multirow{2}{*}{11 (0)}& \multirow{2}{*}{10 (0)}\\
   \cite{harrington_2018}& & & & & \smallskip\\
   \cite{yang_2017, andreani_2018}& \nodata& 7 (4)& \nodata& 11 (0)& 10 (1)\smallskip\\
   \cite{dannerbauer_2019}& 1 (0)& \nodata& 1 (0)& 1 (0)& \nodata\smallskip\\
   \cite{jin_2019}& 1 (0)& \nodata& \nodata& 1 (0)& \nodata\smallskip\\
   \hline
   Total& 35 (8)& 26 (11)& 15 (0)& 59 (0)& 32 (1)\smallskip\\ 
   \enddata
   \tablecomments{Line coverage: $3\sigma$ detections (upper limits).\\
   $^{\dagger}$: The mid- and high-$J$ CO measurements refer only to the \cofour\ and
   \coseven\ that we investigated here. See \cite{liu_2015} for results concerning the remaining
   CO transitions.}
 \end{deluxetable*}

\subsection{Main sequence galaxies}
\label{sec:ms_sample}
This sample is composed of two main sets of observations described in
\cite{valentino_2018} and \cite{bourne_2019}, plus a single object
from \cite{popping_2017}.

\subsubsection{The \cione\ transition}
\label{sec:ci10_cycle4}
$\bullet$ \textbf{\citet[][V18]{valentino_2018}}: In our previous
work, we selected $50$ targets mainly lying on the upper main
sequence at $z\sim1.1-1.3$ in the COSMOS field
\citep{scoville_2007}, while including a subsample of starburst galaxies
(i.e., $>3.5\times$ above the main sequence, Figure 1 in V18). The
targets had available stellar mass estimates
\citep{muzzin_2013, laigle_2016}, a spectroscopic confirmation from
the COSMOS master catalog (M. Salvato et al. in preparation.), and a
\textit{Herschel}/PACS $100$ $\mu$m and/or 160 $\mu$m $3\sigma$
detection in the PEP catalog \citep{lutz_2011}. These galaxies were
followed up in ALMA Band 6 during Cycle 4 (Project ID: 2016.1.01040.S,
PI: F. Valentino), covering \cione\ and, for part of the
sample, \cofour. The ALMA campaign resulted in a secure determination
of $18$ detections and $3$ upper limits on \cione\ down to an average
rms per beam of $\sim0.15$ \jykms\ for a line width of $400$
\kms. We computed the upper limits as $I <
  3\times\mathrm{rms_{\rm ch}}\sqrt{\Delta V dv}$, where
  $\mathrm{rms_{\rm ch}}$ is the average noise per channel over the
  velocity range $\Delta V$ of other securely detected lines for each individual source, and
  $dv$ is the velocity bin size in \kms\ \citep[see, e.g., Eq. 7 of][]{bothwell_2013}.
Here we add two extra sources from that sample with secure
\cione\ detections, but excluded from V18 because of the absence of a
second sub-mm transition to confirm the redshift, now granted by
\citwo\ and \coseven\ (Section \ref{sec:ci21_cycle6}). Furthermore,
all $14$ galaxies with \cofour\ coverage have been
detected during the same runs. Moreover, $15$ and $11$ galaxies
have \cofive\ and \cotwo\ detections as part of independent ALMA programs 
(Project IDs: 2015.1.00260.S, 2016.1.00171.S, PI: Daddi; E. Daddi et
al. 2019, in preparation). A large fraction of this sample (19/23)
is also detected in the 1.1 and/or 3 mm continuum emission. We modeled
the latter together with the whole far-IR SEDs listed in the
``super-deblended'' COSMOS catalog \citep{jin_2018} following the
prescriptions of \cite{magdis_2012}. We adopted the expanded
\cite{draine_2007} models and incorporated the AGN templates by
\cite{mullaney_2011} to derive and subtract the contribution of dusty
tori to the integrated 8-1000 $\mu$m IR luminosity, \lir, 
for every source in the sample. Moreover, we flag as ``AGN''
  objects with at least a 1/3 contribution to the total \lir\ from an
  active nucleus, whose $L_{\rm IR, AGN}$ is detected with
  $S/N>10$. We cross-checked this selection against the IRAC color
  criterion by \cite{donley_2012}, retrieving consistent results. We
  note that our decomposition is sensitive to the coverage
  of the mid-IR wavelength regime and that it is effective to retrieve
  relatively bright AGN. We further remark
that the detection of a millimeter continuum in the Rayleigh-Jeans tail of the
dust emission is critical for a secure determination of the dust mass
\citep[e.g.,][]{magdis_2012, scoville_2014}. Finally, we estimated a
luminosity-weighted dust temperature, $T_{\rm dust}$, by fitting a
modified black body model (MBB) to the SED. We report the line
measurements for this sample in Table \ref{tab:all}. 

  \begin{deluxetable*}{cccccccccccccc}
   \tabletypesize{\scriptsize}
   \tablecolumns{14}
   \tablecaption{Emission lines of main-sequence galaxies at $z\sim1.2$\label{tab:all}.}
   \smallskip
   \tablehead{
     \colhead{ID} & 
     \colhead{$z_{\rm spec}$} & 
     \colhead{\lir}&
     \colhead{$T_{\rm dust}$}&
     \colhead{$\langle U \rangle$}&
     \colhead{\lprimecione}&
     \colhead{\lprimecotwo}&
     \colhead{\lprimecofour}&
     \colhead{\lprimecofive}&
     \colhead{\ici} & 
     \colhead{$I_{\rm{CO(2-1)}}$} &
     \colhead{$I_{\rm{CO(4-3)}}$} &
     \colhead{$I_{\rm{CO(5-4)}}$} &
     \colhead{Type}\\
     \colhead{}&
     \colhead{}&
     \colhead{$10^{11}\,L_{\odot}$}&
     \colhead{K}&
     \colhead{}&
     \colhead{Units}&
     \colhead{Units}&
     \colhead{Units}&
     \colhead{Units}&
     \colhead{\jykms}& 	
     \colhead{\jykms}&
     \colhead{\jykms}& 	
     \colhead{\jykms}\\
     \colhead{(1)}&
     \colhead{(2)}& 	
     \colhead{(3)}&
     \colhead{(4)}&
     \colhead{(5)}&
     \colhead{(6)}&
     \colhead{(7)}&
     \colhead{(8)}&
     \colhead{(9)}&
     \colhead{(10)}&
     \colhead{(11)}&
     \colhead{(12)}&
     \colhead{(13)}&
     \colhead{(14)}
   }
   \rotate
   \startdata
           4233&   $1.1630\pm0.0003$&   $8.09\pm0.49$&   $27.8\pm2.0$&  $9.5\pm2.5$&   $0.24\pm0.05$&   \nodata&   $0.66\pm0.07$&   $<0.21$&   $0.60\pm0.14$&   \nodata&   $1.48\pm0.15$&   $<0.74$&   MS \\
           7540&   $1.1714\pm0.0003$&   $9.36\pm1.32$&   $29.2\pm1.4$&  $8.5\pm1.6$&   $0.36\pm0.08$&   \nodata&   $0.65\pm0.09$&   \nodata&   $0.89\pm0.20$&   \nodata&   $1.42\pm0.19$&   \nodata&   MS \\
          13205&   $1.2660\pm0.0004$&   $12.28\pm1.27$&   $40.0\pm0.9$&  $37.4\pm5.0$&  $<0.21$&   $2.10\pm0.42$&   \nodata&   $1.02\pm0.09$&   $<0.45$&   $0.99\pm0.20$&   \nodata&   $3.03\pm0.27$&   MS \\
          13250&   $1.1484\pm0.0002$&   $5.22\pm1.54$&   $31.0\pm5.0$&  $14.2\pm6.4$&  $<0.10$&   \nodata&   $0.33\pm0.04$&   $<0.10$&   $<0.27$&   \nodata&   $0.74\pm0.10$&   $<0.35$&   MS \\
          18538&   $1.2696\pm0.0001$&   $9.95\pm0.62$&   $34.2\pm1.0$&  $20.5\pm3.4$&  $0.23\pm0.03$&   \nodata&   \nodata&   $0.33\pm0.05$&   $0.49\pm0.07$&   \nodata&   \nodata&   $0.96\pm0.14$&   MS \\
          18911&   $1.1709\pm0.0003$&   $7.78\pm1.68$&   $40.4\pm1.0$&  $51.1\pm7.1$&  $<0.20$ &   $0.81\pm0.26$&   $0.72\pm0.18$&   $<0.43$&    $<0.51$ &   $0.45\pm0.14$&   $1.59\pm0.40$&   $<1.47$&   MS \\
          19021&   $1.2581\pm0.0003$&   $19.27\pm1.57$&   $36.2\pm0.7$&  $23.9\pm1.9$&  $0.48\pm0.09$&   $1.90\pm0.24$&   \nodata&   $0.58\pm0.03$&   $1.06\pm0.19$&   $0.91\pm0.11$&   \nodata&   $1.73\pm0.10$&   AGN \\
          26925&   $1.1671\pm0.0003$&   $9.13\pm1.66$&   $32.2\pm1.3$&  $15.0\pm3.6$&  $0.37\pm0.07$&   $1.47\pm0.22$&   $0.71\pm0.08$&   $<0.20$&   $0.93\pm0.17$&   $0.81\pm0.12$&   $1.58\pm0.17$&   $<0.69$&   MS \\
          30694&   $1.1606\pm0.0002$&   $8.67\pm1.45$&   $33.2\pm5.7$&  $15.0\pm4.6$&  $0.24\pm0.03$&   $1.62\pm0.21$&   $0.48\pm0.04$&   $0.30\pm0.04$&   $0.61\pm0.09$&   $0.91\pm0.12$&   $1.07\pm0.10$&   $1.06\pm0.14$&   MS \\
          32394&   $1.1345\pm0.0001$&   $22.54\pm6.37$&   $33.0\pm1.8$&  $20.5\pm4.6$&  $0.24\pm0.05$&   \nodata&   $0.58\pm0.05$&   \nodata&   $0.64\pm0.12$&   \nodata&   $1.37\pm0.11$&   \nodata&   SB \\
          35349&   $1.2543\pm0.0002$&   $11.50\pm1.83$&   $27.6\pm1.7$&  $6.7\pm2.5$&   $0.41\pm0.11$&   $3.57\pm0.44$&   \nodata&   $0.45\pm0.11$&   $0.89\pm0.25$&   $1.72\pm0.21$&   \nodata&   $1.35\pm0.33$&   MS \\
          36053&   $1.1573\pm0.0003$&   $7.99\pm0.58$&   $32.4\pm1.4$&  $17.0\pm3.7$&  $0.17\pm0.05$&   \nodata&   $0.44\pm0.06$&   $<0.17$&   $0.43\pm0.13$&   \nodata&   $1.00\pm0.13$&   $<0.60$&   MS \\
          36945&   $1.1569\pm0.0003$&   $0.44\pm0.03$&   $36.8\pm2.3$&  $0.9\pm0.1$&   $0.17\pm0.05$&   $0.65\pm0.26$&   $0.35\pm0.06$&   $0.17\pm0.05$&   $0.44\pm0.14$&   $0.37\pm0.15$&   $0.79\pm0.14$&   $0.61\pm0.16$&   AGN \\
          37250&   $1.1526\pm0.0002$&   $16.64\pm0.62$&   $30.4\pm0.6$&  $12.1\pm1.2$&  $0.68\pm0.06$&   $4.58\pm0.32$&   \nodata&   $0.84\pm0.08$&   $1.76\pm0.16$&   $2.60\pm0.18$&   \nodata&   $2.96\pm0.30$&   MS \\
          37508&   $1.3020\pm0.0003$&   $11.20\pm2.17$&   $54.2\pm2.3$&  $81.2\pm5.5$&  $0.31\pm0.08$&   $0.49\pm0.32$&   \nodata&   $0.38\pm0.06$&   $0.64\pm0.16$&   $0.22\pm0.14$&   \nodata&   $1.06\pm0.18$&   MS \\
          38053&   $1.1562\pm0.0003$&   $12.90\pm4.03$&   $37.2\pm1.3$&  $30.0\pm5.8$&  $0.27\pm0.06$&   $1.69\pm0.31$&   $0.60\pm0.06$&   $0.35\pm0.08$&   $0.69\pm0.15$&   $0.96\pm0.17$&   $1.34\pm0.15$&   $1.23\pm0.27$&   SB \\
          44641&   $1.1495\pm0.0002$&   $9.19\pm1.99$&   $30.0\pm0.7$&  $10.0\pm1.1$&  $0.48\pm0.07$&   $1.63\pm0.35$&   $0.87\pm0.08$&   $0.27\pm0.06$&   $1.24\pm0.17$&   $0.93\pm0.20$&   $1.99\pm0.17$&   $0.95\pm0.23$&   MS \\
         121546&   $1.1392\pm0.0002$&   $6.89\pm0.94$&   $32.8\pm1.9$&  $18.0\pm4.9$&   $0.47\pm0.09$ &   \nodata&   $0.87\pm0.15$&   \nodata&    $1.24\pm0.23$ &   \nodata&   $2.01\pm0.34$&   \nodata&   MS \\
         188090&   $1.2364\pm0.0003$&   $28.97\pm1.22$&   $38.8\pm0.4$&  $25.0\pm2.4$&  $0.69\pm0.13$&   \nodata&   \nodata&   \nodata&   $1.56\pm0.29$&   \nodata&   \nodata&   \nodata&   SB \\
         192337&   $1.2884\pm0.0003$&   $9.06\pm0.64$&   $33.2\pm1.0$&  $15.0\pm2.2$&  $0.30\pm0.07$&   \nodata&   \nodata&   \nodata&   $0.62\pm0.14$&   \nodata&   \nodata&   \nodata&   MS \\
         208273&   $1.2662\pm0.0003$&   $4.25\pm0.80$&   $40.8\pm8.5$&  $50.0\pm17.1$& $0.32\pm0.07$&   \nodata&   \nodata&   \nodata&   $0.69\pm0.16$&   \nodata&   \nodata&   \nodata&   MS \\
         218445&   $1.1199\pm0.0004$&   $3.98\pm1.07$&   $24.8\pm2.2$&  $4.3\pm2.4$&   $0.49\pm0.09$&   \nodata&   $0.43\pm0.09$&   \nodata&   $1.33\pm0.25$&   \nodata&   $1.03\pm0.22$&   \nodata&   MS \\
         256703&   $1.2740\pm0.0002$&   $7.05\pm0.92$&   $38.0\pm3.0$&  $32.8\pm10.4$& $0.25\pm0.06$&   \nodata&   \nodata&   \nodata&   $0.53\pm0.12$&   \nodata&   \nodata&   \nodata&   MS \\
   \enddata
   \tablecomments{Column 1: ID. Column 2: spectroscopic redshift. Column 3: total
     IR luminosity integrated within $8-1000$~$\mu$m. Column 4: dust
     temperature. Column 5: dust mass-weighted mean intensity of the radiation field for the
     \cite{draine_2007} models. 
     Columns 6 to 9: galaxy-integrated \lprimecione, \lprimecotwo, \lprimecofour, and
     \lprimecofive. Units: $10^{10}\,$~\kkmspc. Columns 10 to 13: 
     velocity integrated \cione, \cotwo, \cofour, and \cofive\ fluxes.
     Column 14: galaxy type: MS = main-sequence; SB = starburst
     ($>3.5\times$ above the main-sequence); AGN = SED 
     contaminated by
     torus emission.\\
     Upper limits at $3\sigma$. \\
     See \cite{valentino_2018} and Section \ref{sec:ci10_cycle4} for details.\\ 
     (The data are available in the .fits files described in Table \ref{tab:app:columns}.)}
 \end{deluxetable*}

\noindent
$\bullet$ \textbf{\cite{bourne_2019}} presented a set of 10 main sequence galaxies,
selected from the Ultra Deep Survey (UDS) and in the COSMOS
fields based on a SCUBA2 $450$ $\mu$m detection in the SCUBA2 Cosmology
Legacy Survey \citep[S2CLS;][]{geach_2017}. The sample covers the
redshift range $z\sim0.9-1.3$ as determined by the available
\textit{Hubble Space Telescope}/WFC3 G141 grism spectroscopy
\citep{momcheva_2016}. All galaxies have a stellar mass determination
\citep{skelton_2014} and they have been followed up in ALMA band
6 during Cycle 4 and 5 (Project IDs: 2016.1.01184.S and 2017.A.00013.S
PI: N. Bourne). The observations resulted in the
detection of \cione\ at $>3\sigma$ in 6/10 galaxies, a marginal
measurement at $2<\sigma<3$ in 3/10 and an upper limit on 1/10
sources. \cofour\ measurements at
$>4\sigma$ are reported for 4/9 galaxies with proper physical
coverage, along with 2/9 marginal detections at $2<\sigma<3$, and 3
upper limits. 
Continuum emission at $\sim1.1$ mm is detected for 7/10
galaxies. In order to avoid systematics on \lir\ and the dust mass, \mdust, purely
due to modeling, we refitted the deblended far-IR SED
\citep{bourne_2017, bourne_2019} using the same prescriptions reported
in the previous paragraph. This resulted in $\sim0.2$ dex larger
\mdust\ and $<0.1$ dex larger \lir\ than originally listed in
\cite{bourne_2019}, consistently with well known systematics
\citep[][V18]{magdis_2012}.\\

\noindent
$\bullet$ \textbf{\cite{popping_2017, talia_2018}:} Finally, we included the compact main sequence
galaxy GS30274 at $z=2.225$
reported in \cite{popping_2017} and subsequently followed up by \cite{talia_2018}. This object has been selected in
GOODS-South following criteria comparable with the ones in V18
(spectroscopic confirmation, detection in \textit{Herschel}/PACS and
SPIRE), but further requiring ``compactness''
\citep{vandokkum_2015}. This extra criterion results in starburst-like
behavior of some properties \citep{popping_2017, gomez-guijarro_2019b}, which put this object in a likely transitioning
phase. GS30274 has been followed up in ALMA Bands 3 and 4 in Cycle 3 (Project
ID: 2015.1.00228.S, PI: G. Popping), resulting in $>10\sigma$
detections of \cione\, \cothree, and \cofour, along with a $>6\sigma$
continuum emission at $2$ mm. \cite{talia_2018} further reported Band
3 and 6 observations (Project ID: 2015.1.01379.S, PI: P. Cassata) and
a $>9\sigma$ detection of CO(6-5) and the underlying 1.4 mm continuum
at $>5.5\sigma$ significance.
Also in this case, we refit the SED following
\cite{magdis_2012}, retrieving a $\sim25$\% AGN contribution to \lir,
consistent with the observed red \textit{IRAC} colors
\citep{donley_2012} and the results in \cite{talia_2018}. Correcting for the effect of the dusty torus, we
find a $7$\% lower \lir\ and a $2\times$ larger \mdust\ than in \cite{popping_2017}.\\

\noindent
Altogether, we compiled 27 main sequence galaxies at $z\sim0.9-2.2$
detected at $>3\sigma$ in \cione, plus 7 marginal detections or upper
limits. Moreover, 24/34 sources have at least one detection of a 
mid-$J$ CO transition ($J_{\rm  upper}=3-5$), the rest of the sample
having marginal measurements or upper limit on \cofour\ (5/34),
\cofive\ (5/34) or not being covered at the relevant frequency ranges
(1/34). 

 \begin{deluxetable*}{cccccc}
   \tabletypesize{\normalsize}
   \tablecolumns{6}
   \tablecaption{ALMA Band 7 observations of main-sequence galaxies at $z\sim1.2$\label{tab:ci21}.}
   \smallskip
   \tablehead{
     \colhead{ID} & 
     \colhead{\lprimecitwo}&
     \colhead{$L'_{\rm CO(7-6)}$}&
     \colhead{\icitwo} & 
     \colhead{$I_{\rm{CO(7-6)}}$} &
     \colhead{$S_{\rm{850}}$}\footnotesize\\
     \colhead{}&
     \colhead{$10^{10}\,$~\kkmspc}&
     \colhead{$10^{10}\,$~\kkmspc}&
     \colhead{\jykms}& 	
     \colhead{\jykms}&
     \colhead{mJy}\footnotesize\\
     \colhead{(1)}&
     \colhead{(2)}& 	
     \colhead{(3)}&
     \colhead{(4)}&
     \colhead{(5)}&
     \colhead{(6)}
   }
   \startdata
          18538&    $0.11\pm0.01$&   $0.10\pm0.01$&   $0.63\pm0.07$&   $0.57\pm0.07$&   $1.20\pm0.08$ \\
          19021&    $0.25\pm0.03$&   $0.28\pm0.02$&   $1.45\pm0.16$&   $1.62\pm0.14$&   $2.38\pm0.10$ \\
          35349&    $0.09\pm0.01$&   $0.07\pm0.01$&   $0.53\pm0.09$&   $0.40\pm0.07$&   $1.86\pm0.06$ \\
         188090&   $0.38\pm0.06$&   $0.31\pm0.06$&   $2.29\pm0.35$&   $1.90\pm0.35$&   $3.50\pm0.15$ \\
         192337&   $0.14\pm0.02$&   $0.13\pm0.02$&   $0.80\pm0.10$&   $0.73\pm0.11$&   $1.16\pm0.05$ \\
         208273&   $0.09\pm0.01$&   $0.07\pm0.01$&   $0.54\pm0.07$&   $0.41\pm0.07$&   $0.71\pm0.04$ \\
         256703&   $0.15\pm0.01$&   $0.09\pm0.01$&   $0.87\pm0.09$&   $0.51\pm0.08$&   $1.16\pm0.05$ \\
  \enddata
   \tablecomments{See Table \ref{tab:all} for additional available
     transitions. Column 1: ID. Columns 2 and 3:
     galaxy-integrated \lprimecitwo\ and \lprimecoseven\ luminosities.
     Columns 4 and 5: 
     velocity integrated \citwo\ and \coseven\ fluxes. Column 6: continuum
     emission flux density at $850$ $\mu$m.\\
     (The data are available in the .fits files described in Table \ref{tab:app:columns}.)}
 \end{deluxetable*}

\subsubsection{The \citwo\ transition}
\label{sec:ci21_cycle6}
During ALMA Cycle 6, we collected Band 7 observations for a set of 7
galaxies extracted from the sample in V18 (Project ID:
2018.1.00635.S, PI: F. Valentino). We selected the targets based on
a secure \cione\ detection, the simultaneous observability of \citwo\
and \coseven, the availability of alternative line
emissions (\cofive+\cotwo\ for 2/7 objects, \cofive\ only for 1/7), and a well
constrained IR SED, allowing us to derive dust masses and total
\lir\ (Section \ref{sec:ci10_cycle4}). We selected
  5 typical main sequence galaxies, 1 starburst, and 1
  AGN. We targeted the \citwo\ and
\coseven\ lines within contiguous spectral windows of 1.875 GHz and
with a spectral resolution of 7.8 MHz ($\sim7$~\kms), enough to
spectrally resolve the emission lines. Five out of $7$ targets were
observed for the full proposed integration, the remaining being imaged
for 75\% of the initial request, resulting in a higher rms (\# 35349
and 208273 in Table \ref{tab:ci21}). Data were
collected in the C43-1 configuration for a final synthesized beam of
$\sim0.9$''. We resolved the emission of every source, ensuring
minimal flux losses with Gaussian extractions 
(Appendix \ref{sec:app:totflux}, \citealt{coogan_2018,
  puglisi_2019}). The data were reduced with a
combination of the standard pipeline with CASA \citep{mcmullin_2007}
and a series of customized scripts with GILDAS\footnote{\url{http://www.iram.fr/IRAMFR/GILDAS}}
\citep{guilloteau_2000}, following the procedure described in V18 and
\cite{daddi_2015}. We consistently extracted 1D spectra for all the
available lines, centering on the
brightest peaks among all the transitions available. We scanned the 1D
spectra and integrated the line fluxes over the number of channels
maximizing the S/N ratio of each candidate line emission. These line fluxes
were then increased by 10\% to account for emission in
extended wings as found by modeling the spectra with single or double
Gaussian peaks. We performed such modeling using the
$\chi^2$-minimization algorithm \textsc{MPFIT} \citep{markwardt_2009} and using
both single and double Gaussians with constant velocity widths. In 6/7
cases the line emission are well fitted by double-peaked \citwo\ and \coseven\
profiles. This resulted in a 100\% detection rate of both
transitions at $\gtrsim6\sigma$. We measured line ratios by fixing the
redshifts and line widths to the
values for the highest S/N transitions among the ones available for
each source, with the exception of \#35349, for which the \citwo\ and
\coseven\ lines are wider than \cione\ and \cotwo. For the vast majority of our sources, the estimates are
fully consistent with the ones reported in V18. In a few cases
(notably \#188090 and \#35349) the results significantly varied based on the new
\citwo\ and \coseven\ broad line detections.
We concurrently measured the continuum emission at observed
$\sim850$~$\mu$m over 7.5 GHz assuming an intrinsic slope of $\nu=3.7$
($\beta = 1.7$), excluding the channels covered by the
emission lines. We detected significant continuum emission at
$15-30\sigma$ in 7/7 sources.\\ 
 
All the line measurements and the underlying $850$~$\mu$m continuum
emission are reported in Table \ref{tab:ci21}.

\subsection{Local galaxies}
\label{sec:local_sample}

$\bullet$ \textbf{\cite{liu_2015}:} This sample is composed of 
galaxies from a compilation of \textit{Herschel}/Fourier Transform Spectrometer
(FTS) observations in the Herschel Science Archive of local
galaxies. We retrieve 32 (126) objects with a \cione\  (\citwo)
detection at $>3\sigma$. All 32 sources with a \cione\ measurement are detected in
\citwo. Multiple $J_{\rm upper}>4$ CO lines are generally
available \citep{liu_2015}. In particular, 31/32 sources with \cione\
and 105/126 objects with \citwo\ have coverage of the \cofour\ line
(55 detections). All sources have coverage of \coseven\ (104 detections).
For consistency, we checked our beam flux
measurements against the independent analysis of
\cite{kamenetzky_2016} and \cite{israel_2015}, recovering consistent
results for the sources in common among these samples.
We corrected the $L_{\rm FIR}(\mathrm{40-400\,\mu m})$ luminosities from \textit{IRAS}
\cite{sanders_2003} to \lir\ by multiplying by a factor of
$1.2\times$. This average correction was checked against full SED
modeling for a subset of galaxies
from the Great Observatories All-Sky LIRGs Survey
\citep[GOALS;][]{armus_2009}. For such subsample, we further estimated
the dust temperature $T_{\rm dust}$ by fitting an MBB model as for the main sequence galaxies.   
As described in V18, we beam-matched
the line luminosities to \lir\ based on \textit{Herschel}/PACS
photometry. Therefore, the values adopted in our analysis refer to the
total, galaxy-integrated quantities. We further include and beam-match
the observations of low-$J$ CO transitions from
\cite{kamenetzky_2016}. We finally checked for signatures of
galaxy nuclear activity by cross-matching our sample with the catalog
by \cite{veron-cetty_2010}, retrieving 12/32 and 43/126
galaxies that we therefore flag as ``active''. Given the observed
luminosities and properties, the local galaxy sample is representative
of the starbursting population, rather than typical low-redshift spirals
(V18).

\subsection{High-redshift Submillimeter Galaxies and Quasars}
\label{sec:smg_sample}
We collected information about recent observations of the \cione\
and/or \citwo\ transitions in high-redshift SMGs and QSOs. For these
objects, we retrieved the original far-IR to
sub-mm SED and refitted it following the same procedure and adopting
the identical models as for the main sequence galaxies in V18
described in the previous section. As noted in
that paper, this general results in $\sim1.5\times$ larger 8-1000
$\mu$m \lir\ and up to $10\times$ larger \mdust\ than the widely
adopted MBB law \citep[e.g.,][]{blain_2003,
  magdis_2012, dale_2012, bianchi_2013}. The
difference is larger for sources particularly bright in the
mid-IR (e.g., AGN/QSOs), where the difference between the MBB
and the \cite{draine_2007} models is maximal. For the same reason,
the more divergent the integration limits of the
``far-IR'' luminosities (\lfir, 40-120 or 40-400 $\mu$m,
depending on the convention) from the total \lir\ (8-1000 $\mu$m), the
greater the correction to apply. These differences are well known and
entirely due to the adopted models and their parameters (effective dust emissivity
index $\beta$, dust mass absorption coefficient $\kappa$, peak temperature;
\citealt{magdis_2012}). Only by correcting for these \textit{systematic}
deviations, we can safely compare the \textit{relative} behavior of the various
galaxy populations. We also notice that the vast majority of
galaxies in this sample does not have an estimate of the stellar mass
and we cannot canonically define them as main sequence or starburst
galaxies. However, their observed ISM conditions, gas and SFR
densities, and SFEs generally distinguish SMGs from main galaxies, and
we will thus consider them as starbursts as in our previous
analysis (V18).\\

\noindent
Altogether we collected information about 60 SMGs at $z\sim2-5$,
35/60 detected in \cione\ (8/60 upper limits) and
26/60 detected in \citwo\ (11/60 upper limits, Table \ref{tab:compilation}). Moreover, 59/60
sources have at least one detection of a 
mid-$J$ CO transition ($J_{\rm  upper}=3-5$) and 21/60 are detected in
\coseven\ (4/60 upper limits).\\ 

\noindent
$\bullet$ \textbf{\cite{walter_2011}, \cite{alaghband-zadeh_2013}:} 
These authors targeted or collected information on typical $850$
$\mu$m selected SMGs at $z\sim2.5-4$, including a high-redshift tail
of widely known and studied QSOs. Half of the sample is gravitationally
magnified up to $\sim30\times$ and 30\% is contaminated or dominated
by the emission of dusty tori surrounding the central supermassive
black hole.
Out of 23 galaxies, 17 are detected in \cione\ and 11 in
\citwo\ (10 galaxies have both lines available). Moreover, 4/23 and
7/23 objects have upper limits on \cione\ and \citwo, respectively. We notably
substituted the old upper limits on the \ci\ transitions in GN20 at
$z=4.055$ \citep{daddi_2009, casey_2009} with the recent detections
with the NOEMA interferometer (Cortzen et al., submitted). The vast
majority of the sample (22/23) has a secure detection of \cofour\
or \cothree\ (5/23 galaxies have both line fluxes
available). Moreover, 12/23 objects have a detection of \coseven\
(1/23 upper limits). Sixty-five percent of these galaxies have
 interferometric observations.\\

\noindent
$\bullet$ \textbf{\cite{bothwell_2017}:} This sample comprises 13
strongly lensed systems ($\mu_{\rm magn}\sim3-30\times$) found in the
1.4 mm blank-field survey with the South Pole Telescope
\citep[SPT;][]{vieira_2010, weiss_2013}, spectroscopically confirmed
to lie at $z=3.3-4.8$ by multiple line transitions, including both
high- and low-$J$ CO transitions \citep{weiss_2013,
  aravena_2016} and ionized carbon emission \ciplus\
\citep{gullberg_2015}. \cite{bothwell_2017} reports \cione\
ALMA detections at
$>3\sigma$ significance for 9/13 galaxies. No coverage of the \citwo\
line is available. Our SED modeling identifies only 1/13 source with a
significant contribution ($f_{\rm AGN}\sim70$\%) of the central AGN to
the total \lir. \\ 
\begin{figure*}
  \centering
  \includegraphics[width=\columnwidth]{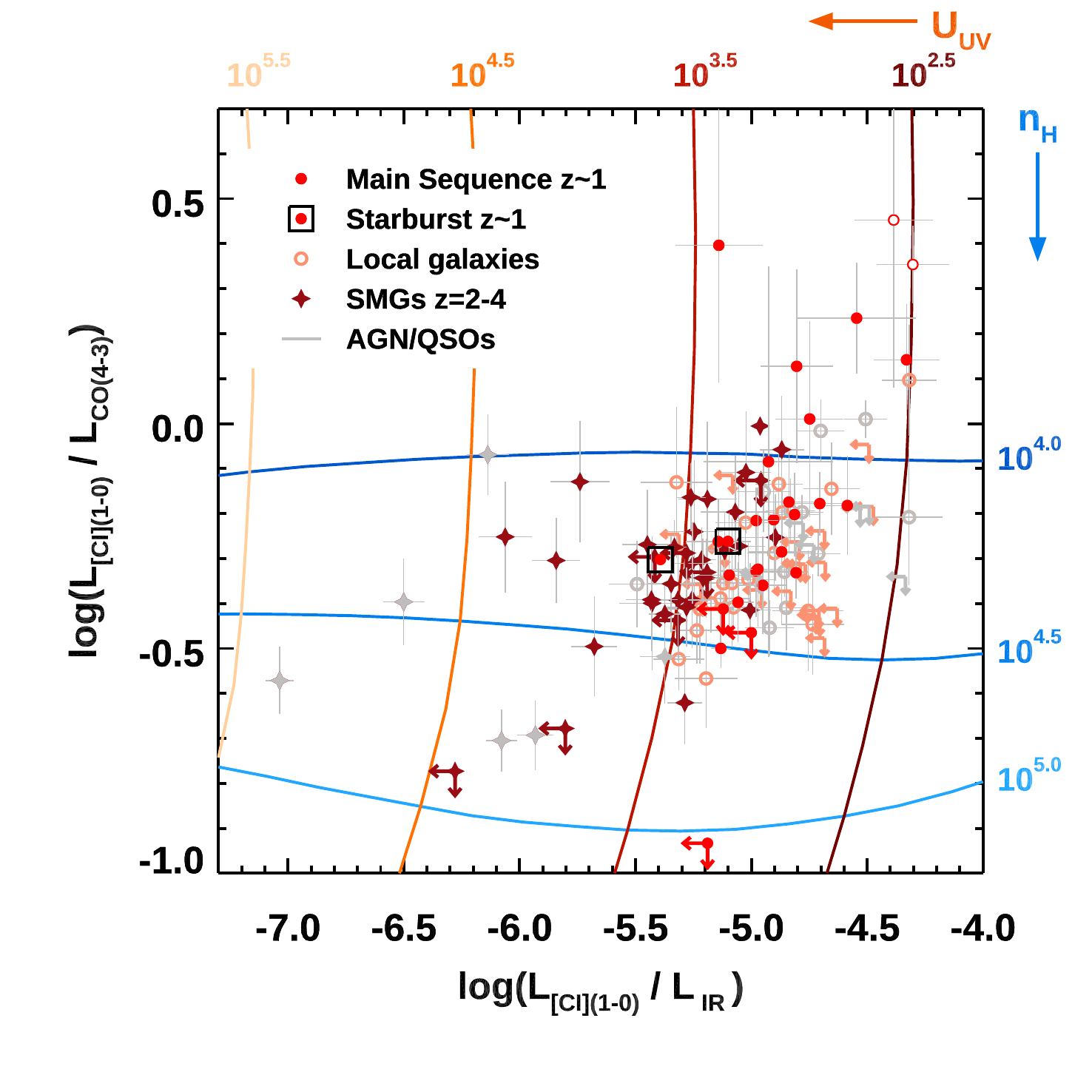}
  \includegraphics[width=\columnwidth]{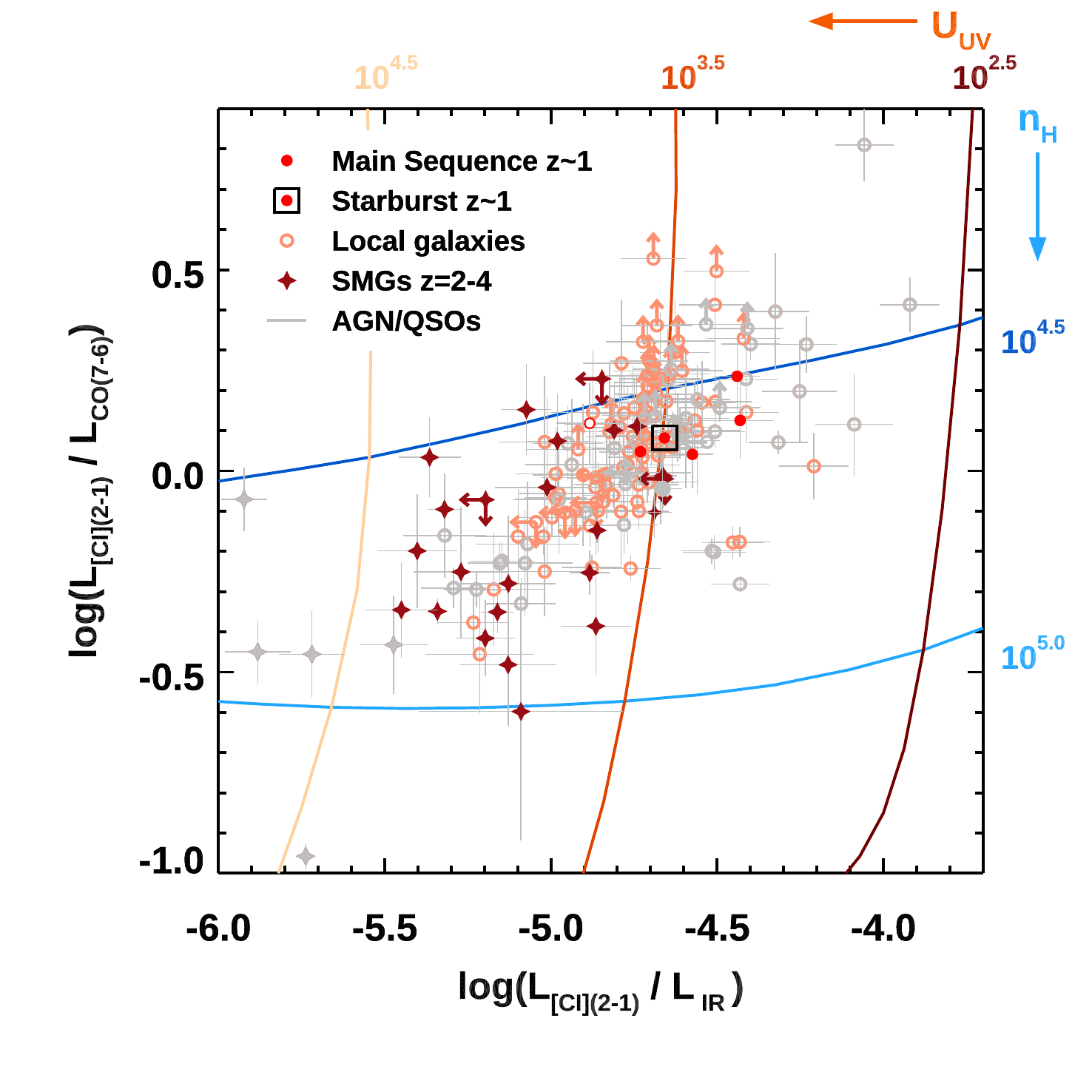}
  \caption{\textbf{\ci, \co, and \lir\ planes.} \textit{Left}:
    \lci/\lir\ -- \lci/$L_{\rm CO(4-3)}$. \textit{Right:} \lcitwo/\lir\ --
    \lcitwo/$L_{\rm CO(7-6)}$. The luminosities
      are in \lsun. Red solid
    circles: main sequence galaxies at $z\sim1$ (this work, V18,
    \citealt{bourne_2019}; red empty circles: sources with caveats from the
    latter) and $z=2.225$ \citep{popping_2017}; empty black squares:
    starbursts at $z\sim1.2$ (V18); empty orange circles:
    local FTS sample of star-forming
    galaxies \citep{liu_2015, kamenetzky_2016}; empty gray circles:
    local FTS sample with AGN signatures \citep{liu_2015,
      kamenetzky_2016, veron-cetty_2010}; dark red solid stars: $z\sim2-4$
    SMGs \citep{walter_2011, alaghband-zadeh_2013, bothwell_2017,
      yang_2017, andreani_2018, canameras_2018, nesvadba_2018,
      dannerbauer_2019, jin_2019}.  Gray symbols indicate
    QSO/AGN-contaminated galaxies. Arrows mark $3\sigma$ limits. 
    When not available, we derived \cofour\ from \cofive\ or
    \cothree\ by applying
    average corrections for each individual sample (Section \ref{sec:lineratios}). The
    blue and red solid lines indicate the tracks of constant density
    $n\,\mathrm{[cm^{-3}]}$ and intensity of the UV radiation field
    $U_{\rm UV}\, [\rm{Habing \; units, G_0}]$ from the PDR models
    by \citet{kaufman_1999}. }
  \label{fig:obsratios} 
\end{figure*}

\noindent
$\bullet$ \textbf{\cite{canameras_2018, nesvadba_2018}:} These authors
report IRAM/EMIR single-dish
observations of \cione\ and/or \citwo\ in a subsample
of 11 galaxies from the \textit{Planck}'s dusty Gravitationally
Enhanced subMillimetre Sources \citep[GEMS;][]{canameras_2015}. These
objects have been initially selected as the brightest among the isolated,
compact sources with the reddest $350-550$ $\mu$m and $550-850$ $\mu$m
\textit{Planck} colors and subsequently followed up with multiple
facilities that sampled their far-IR / sub-mm SED and confirmed
their redshift with several line transitions
($z=2.2-3.5$, including \coone\ from \citealt{harrington_2018}). The
magnification factor is generally well constrained
both for the continuum and the line emission, spanning a range of
$\mu_{\rm magn}=7-30\times$. When necessary, in the analysis we
adopted separate $\mu_{\rm dust}$ and $\mu_{\rm gas}$ to correct the
continuum emission and its derived properties (e.g., \lir, \mdust) and
the line luminosities \citep[from Table 1 of][]{canameras_2018}. All 7/11 and 8/11
galaxies with \cione\ and \citwo\ coverage, respectively, are securely detected
\citep{nesvadba_2018}. Four out of 11 sources have both
transitions available. Our SED modeling confirms the lesser AGN contribution to
the total \lir\ reported in \cite{canameras_2015}
($\mathrm{max}(f_{\rm AGN})\sim30$\% for 3/11 galaxies, negligible for
the rest of the sample).\\

\noindent
$\bullet$ \textbf{\cite{dannerbauer_2019}:} Dubbed the ``Cosmic
Eyebrow'' in analogy with the prototypical strongly lensed SMG
``Cosmic Eyelash'' \citep{ivison_2010b, danielson_2011}, this source
has been selected by cross-matching the
\textit{AllWISE} and the \textit{Planck} full-sky compact source
catalogs \citep{diaz-sanchez_2017}. The red \textit{WISE} colors and
ultrabright emission detected by \textit{Planck} and SCUBA2
\citep{jones_2015} have been recently confirmed to arise from two
lensed galaxies at $z=2.04$ by \cite{dannerbauer_2019}, who report
\cione,  \cofour, \cothree, and \coone\ fluxes measured with NOEMA,
IRAM/EMIR and GBT, respectively. The NOEMA observations
spatially resolve the \cothree\ emission from the A and B components
and allow for the deblending of the far-IR emission, assuming an
average observed (i.e., magnified) luminosity ratio of $2.8\times$
between the two galaxies. We have assumed this value in order to split the global
properties that we derived from the SED modeling (e.g., $L_{\rm IR,
  tot} = \mu_{\rm A}L_{\rm A} + \mu_{\rm B}L_{\rm B} = (1+1/2.8)\,
\mu_{\rm A}L_{\rm A}$, with $\mu_{\rm A}=11\pm2$ and $\mu_{\rm
  B}=15\pm3$). Here we consider only the component A, to
which the \cione\ and \cofour\ line emissions are associated.\\  

\noindent
$\bullet$ \textbf{\cite{yang_2017, andreani_2018}:} These authors
report \citwo\ measurements for 11 galaxies drawn from a 
subsample of SMGs from the \textit{Herschel}-Astrophysical Terahertz
Large Area Survey \citep[\textit{H}-ATLAS;][]{eales_2010}. The sources
have been selected based on their bright \textit{Herschel}/SPIRE $500$
$\mu$m fluxes ($S_{\rm 500}>100$ mJy), a suitable threshold to identify
strongly lensed dusty systems \citep[e.g.,][]{negrello_2010}. The
redshift confirmation at $z=2-3.5$ mainly came
  from \coone\ \citep{harris_2012}, followed by NOEMA, ALMA, and
APEX/SEPIA 5 campaigns detecting several sub-mm transitions, including
both molecular and atomic species \citep[CO, H$_2$O, and
\citwo][]{yang_2017,andreani_2018}. For our analysis, when available we adopted the
photometry in \cite{zhang_2018} and the magnification factors mainly
derived from $880$ $\mu$m observations \citep{bussmann_2013}. When not
available, we used the photometry in \cite{yang_2017}.
In total, we retrieve 7 detections at $>3\sigma$, 2 marginal
detections and 2 upper limits on \citwo. No
coverage of the \cione\ line is available for these sources. All 11
galaxies have at least one detection in \cofour, \cofive, or \coseven. \\

\noindent
$\bullet$ \textbf{\cite{jin_2019}:} We include the SMG at $z=3.623$
(ID:85001674) with \cione\ and \cofour\ detections in 
ALMA Band 3 (Project
ID: 2017.1.00373.S, PI: S. Jin). This and a handful of other sources
were selected as residuals in the COSMOS/SCUBA2 $850$~$\mu$m map, 
after the subtraction of known bright sources \citep{geach_2017}.
Here we adopted their ``intrinsic''
quantities obtained including the effect of the
  CMB. Notice that the authors rely on a MBB law, since the CMB effect
  cannot currently be included in \cite{draine_2007} models. However, the
  final quantities have been corrected to total values to match the
  same conventions we adopt here. The results do not change using the
``observed'' values (see Table 3 in \citealt{jin_2019}).

\begin{figure*}
  \centering
  \includegraphics[width=\textwidth]{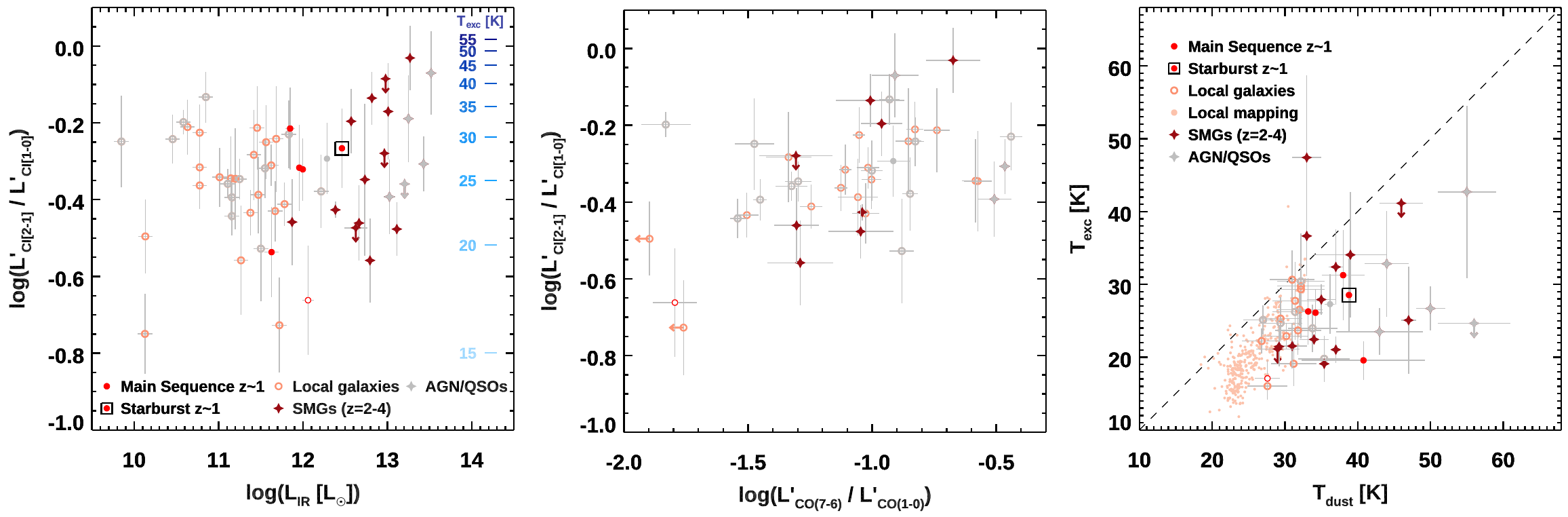}
  \caption{\textbf{\citwo/\cione\ line ratio in
      galaxies.} \textit{Left:} \lprimecitwo/\lprimeci\ ratio as a
    function of the total \lir. Symbols and colors indicate galaxies as in Figure
    \ref{fig:obsratios}, when both \cione\ and \citwo\ detections are
    available. The blue ticks
    show the excitation temperature $T_{\rm exc}$ corresponding to the
  \lprimecitwo/\lprimeci\ ratio on the Y-axis, assuming a local
  thermal equilibrium and optically thin \ci\ lines. \textit{Center:}
  \ci\ line ratio as a function of \lprimecoseven/\lprimecoone. 
  \textit{Right:}
  the excitation temperature $T_{\rm exc}$ as a function of the dust
  temperature from SED fitting with a single-component, optically thin
  modified black body model. The filled small
    orange circles show the resolved mapping from
    \cite{jiao_2019_erratum}, with $T_{\rm dust}$ from
    \textit{Herschel} colors. The dashed black line shows the
  one-to-one relation. For clarity, we do not show the lower limits on
  the \ci\ line ratios and on $T_{\rm exc}$ for the local sample.  
}
  \label{fig:ciobsratios}
\end{figure*}

\section{Analysis and results}
\label{sec:results}

\subsection{\ci\ and mid-/high-$J$ CO line ratios}
\label{sec:lineratios}
In V18 we showed that the \cione/low-$J$ CO luminosity ratio ($J_{\rm upper}=1-2$) is constant
over $z=0-4$ and irrespectively of the galaxy type, suggesting that \ci\ and CO are
well correlated on global scales. This extended previous findings for local
nuclear starbursts \citep[e.g.,][now
confirmed on sub-galactic scales, \citealt{jiao_2019}]{gerin_2000, jiao_2017} and
high-redshift SMGs \citep{yang_2017} to the bulk of main sequence galaxies. Here
we focus on the comparison between \ci\ and mid-/high-$J$ CO emission ($J_{\rm
  upper}=3-7$), the latter expected to arise from warmer and denser
molecular phases. Figure \ref{fig:obsratios} shows \lcione/\lcofour\ as a
function of \lcione/\lir\ and \lcitwo/\lcoseven\ against
\lcitwo/\lir. The luminosities are all expressed in \lsun. 
The choice of these line ratios is dictated by the close rest-frame
frequencies of these \ci-CO couples, which make them often 
observed simultaneously. 
The use of line ratios mitigate the consequences of lensing, but
  a differential effect can still affect the
  emission arising from distinct ISM phases. We refer the
  reader to the original works for a detailed description of
  differential lensing on high-redshift sub-mm galaxies. On the
  contrary, this is not a
  concern for local and main sequence galaxies. 
Since some galaxies do not have
direct observations of the \cofour\ line available, but were observed in a
adjacent transition, we used the following line ratios to correct to
\cofour: \lcothree/\lcofour$=0.57$ and
\lcofive/\lcofour$=1.36$ for a total of 9 high-redshift SMGs
and \lcofive/\lcofour$=0.92$ for 5 main sequence galaxies. 
These factors are the observed average values for galaxies in
  each sample with both CO lines available. The results
do not change adopting median corrections. The slight difference in
\lcofive/\lcofour\ for SMGs and main sequence galaxies 
might suggest a
different slope of the CO SLED \citep[e.g.,][]{daddi_2015},
but it is not significant at this stage. No
correction was applied to \lcitwo/\lcoseven.\\

In both panels we find a gradient of observables across the
populations. High-redshift SMGs appear to have lower \lcione/\lcofour\
and \lcitwo/\lcoseven\ ratios than the main sequence and local
LIRGs. A similar trend is appreciable for \lcione/\lir\ and
\lcitwo/\lir, as commented in V18. We quantify the differences in the 
luminosity ratio parent distributions by running the set of non-parametric two-sample tests
from the \textit{twosampt} task in the \textsc{IRAF/STSDAS} package \citep{feigelson_1985},
including the censored data, generally in the form of upper limits on
the \ci\ luminosities. A notable exception is a substantial number
of lower limits ($21$ galaxies) on the \lcitwo/\lcoseven\ ratios for the sample of local
LIRGs without AGN signatures. Since doubly censored data are not
allowed by the two sample tests, we separately ran the latter on the
population of lower and upper limits. This suite of tests includes the
logrank, the Gehan, Peto \& Peto, and Peto \& Prentice Generalized Wilcoxon tests. 
Moreover, Table \ref{tab:lineratios} reports the mean and its uncertainties
for each sample using the \citet{kaplan-meier_1958} estimator. We
exclude galaxies with a substantial contribution to the IR emission and,
potentially, to the line excitation from AGN/QSOs, since we cannot
securely disentangle the contribution to the dust emission heated by
star formation from the SED modeling. We show the position of such
galaxies in the plots, but their properties are driven by the
large \lir\ ensuing the emission from dusty tori in the mid-IR (V18).\\

The probability that the observed distributions of \lcione/\lir\ ratios are drawn
from the same parent distributions is $p<0.0001$ for all the tests
when comparing the high-redshift SMGs and the main sequence
galaxies at $z\sim1$. Similarly we retrieve  $p<0.006$
 when considering
the SMGs and the local galaxies, with the exception of the Peto \&
Prentice generalized Wilcoxon test returning a $p$-value of 
$p=0.044$.  We obtain marginally consistent distributions when
comparing local and main sequence galaxies at $z\sim1$
 ($p=0.006-0.016$).  Turning to \lcione/\lcofour, we find evidence for
different parent distributions when comparing local and main sequence
galaxies  ($p<0.004$) , but not for main sequence objects and SMGs
 ($p=0.062-0.56$),  nor local galaxies and
SMGs ($p=0.033-0.089$, except
for the Peto \& Prentice test returning $p=0.001$). For what concerns
the distributions of the \lcitwo/\lir\ ratios, we
find $p<0.001$ for all the tests
when comparing the high-redshift SMGs and the local
galaxies. Similarly, we find $p<0.002$ when comparing main sequence
objects at $z\sim1$ and SMGs, with the exception of the Peto \&
Prentice test ($p=0.014$). On the contrary, the local and main sequence
galaxies are consistent with being drawn from the same parent
distributions ($p=0.03-0.49$), the $p$-values spanning a different
range when considering separately upper and lower limits in the local
sample, but still safely larger than meaningful thresholds to reject
the null hypothesis. An identical conclusion is reached when comparing
the \lcitwo/\lcoseven\ ratios ($p=0.02-0.76$ when comparing local and
main sequence galaxies at $z\sim1$; $p=0.0-0.021$ for main sequence
and high-redshift SMGs; $p<0.002$ for local galaxies and SMGs). We
stress that only a handful of main sequence galaxies with \citwo\ and \coseven\
measurements are currently available and, thus, these results will have to
be validated with larger samples.\\

The results of the tests on the \ci/IR luminosity ratios confirm what
we found in the less numerous sample of V18: the local starbursts and
$z\sim1$ main
sequence galaxies share similar
properties, while significantly differing from the high-redshift
SMGs. Here we reach similar conclusions also for the
\lcitwo/\lcoseven\ ratios, pointing towards an intrinsic difference
of the physical properties of the dense and diffuse gas in these
populations. The conclusions based on \lcione/\lcofour\ are less
clear: our samples appear more homogeneous, as noted for \cione/\cotwo\
in V18. This might be due to the CO SLEDs
  of the different samples being more similar at low and mid-$J$
  transitions, while clearly diverging at higher $J_{\rm upper}$, where
  the distributions keep rising for strongly starbursting systems,
  while declining for ``normal'' disks \citep[e.g., ][]{liu_2015,
    daddi_2015, yang_2017, canameras_2018}.\\

Since \ci, mid-/high-$J$ CO, and
\lir\ trace the low density, high density gas, and the SFR,
respectively, we interpret these trends as
evidence for increased dense molecular gas fractions and higher star formation efficiencies
($\mathrm{SFE} = \mathrm{SFR}/M_\star$) in high-redshift SMGs than
main sequence galaxies and local LIRGs. Assuming that the SMG population is
dominated by starbursting galaxies (i.e., several times above the main
sequence at their redshifts), this is consistent with the picture
derived from classical CO-based studies
\citep[e.g.,][]{solomon_1997, gao_1999, gao_2004,
  daddi_2010b,genzel_2015,yamashita_2017,tacconi_2018} and dense molecular gas
tracers \citep[e.g., HCN,][]{gao_2007}. \coseven\ is
arguably a better tracer of the dense
molecular gas than mid-$J$ ($J_{\rm upper}=3-5$) and, thus, the
separation of the various populations in the right panel of Figure
\ref{fig:obsratios} is more evident (despite the general caveat of possible
contributions of XDR, \citealt{meijerink_2007}, or shocks, \citealt{lee_2019},
to the high-$J$ CO emission). Alternatively, these panels might
be understood as manifestations of the (integrated) Schmidt-Kennicutt
relation \citep{schmidt_1959, kennicutt_1998b}. Again, the better
separation of the various populations in the right panel of Figure \ref{fig:obsratios} derives from
the tighter correlation with \lir\ of \coseven\ than \cofour\ \citep{greve_2014,liu_2015,
  kamenetzky_2016, lu_2015, lu_2017}. \coseven\ shows in
fact the tightest correlation with \lir\ among the CO transitions, at
least in the local universe \citep[e.g.,][]{lu_2015, liu_2015}.

 \begin{deluxetable*}{lccc}
   \tabletypesize{\normalsize}
   \tablecolumns{4}
   \tablecaption{Mean line luminosity ratios.\label{tab:lineratios}}
   \smallskip
   \tablehead{
     \colhead{}&
     \colhead{log(\lcione/\lir)}&
     \colhead{log(\lcione/\lcofour)}&
     \colhead{Detections/Censored}
     \smallskip
   }
   \startdata
   Local galaxies  & $-5.075 \pm 0.043$ & $-0.390 \pm   0.027$ & $19 / 18$ \\
   Main sequence $z\sim1$ &  $-4.908 \pm  0.044^\dagger$
   &  $-0.243 \pm  0.064^\dagger$  & $21 / 3$\\
   SMGs $z\sim2-4$ &  $-5.336 \pm  0.060^\dagger$  & $-0.348  \pm  0.032^\dagger$ & $29 / 7$\\
   \hline
   \rule{0pt}{3ex} & log(\lcitwo/\lir) & log(\lcitwo/\lcoseven) & \smallskip\\
   \hline
   Local galaxies (upper) & $-4.792 \pm  0.025$ & $-0.011 \pm  0.020$ & $62/5$\\ 
   Local galaxies (lower) & $-4.707 \pm  0.025$ & $0.108 \pm  0.027^\dagger$ & $62/21$\\ 
   Main sequence $z\sim1$ & $-4.612 \pm  0.078$  & $0.114 \pm  0.031$ & $5/0$\\ 
   SMGs $z\sim2-4$ & $-5.084 \pm  0.050$ & $-0.193 \pm  0.044$ & $21 / 3$\\
   \hline
   \rule{0pt}{3ex} &  \multicolumn{2}{c}{\lprimecitwo/\lprimecione} & \smallskip\\
   \hline
   Local galaxies & \multicolumn{2}{c}{$ 0.437 \pm  0.028$} & $20/0$\\ 
   Main sequence $z\sim1$ & \multicolumn{2}{c}{$0.465 \pm  0.057 $}  & $4/0$\\ 
   SMGs $z\sim2-4$ &\multicolumn{2}{c}{$0.476 \pm  0.057 $}  & $10 / 3$\\
   \enddata
   \tablecomments{ $^\dagger$: The mean value is formally biased, since the lowest
     value is an upper limit.}
 \end{deluxetable*}

\subsection{\ci\ line ratios and gas temperature}
\label{sec:ciratio}
Given its simple three-level structure, the \citwo/\cione\ line ratio
can serve as a measurement of the gas kinetic temperature (see
\citealt{papadopoulos_2004} for a full derivation). Under the
assumption of local thermal equilibrium, the kinetic temperature
equals the excitation temperature $T_{\rm exc} / \mathrm{K} = 38.8 /
\mathrm{ln}(2.11/R)$, where $R =
L'_{\mathrm{[C\,\scriptscriptstyle{I}\scriptstyle{]}}^3P_2\,-\, ^3P_1}
/ L'_{\mathrm{[C\,\scriptscriptstyle{I}\scriptstyle{]}}^3P_1\,-\,
  ^3P_0}$, which further requires the lines to be optically thin
\citep{schneider_2003, weiss_2003, walter_2011}.\\

We show in Figure \ref{fig:ciobsratios} the available galaxies
with both \ci\ lines. The observed \lprimecitwo/\lprimecione\ ratios
of the local, main sequence at $z\sim1$ and
high-redshift SMG samples are fully consistent (Table
\ref{tab:lineratios}) and we do not find any significant correlation
with \lir. Notice that we considered only the detections
of both \ci\ lines for the local sample, since the large number of
upper limits on \cione\ does not affect the mean value reported
here. We further excluded \# 35349 from the sample of main sequence
galaxies, given the mismatch of the two \ci\ line profiles. 
Converting the line ratios into temperatures, we find a mean temperature
of $\langle T_{\rm exc} \rangle = 25.6 \pm 1.0$~K for the whole compilation, including
upper limits. Nevertheless, the scatter of the measurements is
substantial: we find an interquartile range of $\Delta T_{\rm exc} =
8.0$~K centered on a median value of $T_{\rm exc} = 25.1$~K.
These values are consistent with our estimates in V18 and with
previous measurements reported for individual subsamples ($\langle
T_{\rm exc} \rangle=29.1\pm 6.3$~K, \citealt{walter_2011}; $T_{\rm
  exc}=21-57$~K, \citealt{nesvadba_2018}). Note that we excluded
the objects contaminated by AGN/QSOs from this calculation. The mean
temperature value is slightly lower than the commonly adopted $T_{\rm
  exc}=30$~K \citep[e.g.,][]{alaghband-zadeh_2013,
  bothwell_2017}. While this has a minor impact on the calculation of
\ci\ masses from \cione, lower temperatures affect such estimates
using \citwo\ as a tracer \citep[e.g., Figure 2 in][]{weiss_2005}.\\

We further compared the excitation temperature $T_{\rm exc}$
($\propto$~\lprimecitwo/\lprimecione) with the luminosity-weighted dust temperature
$T_{\rm dust}$ from the modeling of
the SED with a single-component, optically thin modified black body curve 
(Figure \ref{fig:ciobsratios}). We chose this
  simple parametrization to facilitate the comparison with literature
  data. However, similar conclusions can be drawn when
comparing $T_{\rm exc}$ with the mean
intensity of the radiation field $\langle U \rangle$ from the
multi-component
\citet{draine_2007} models, an alternative tracer of the dust heating
correlated with a mass-weighted $T_{\rm dust}$ ($\langle U
\rangle = (T_{\rm dust}[\mathrm{K}]/18.9)^{6.04}$, \citealt{magdis_2017,schreiber_2018_dust}). 
The dust temperature is frequently assumed
as a proxy for $T_{\rm exc}$ and the gas temperature, absent a direct
estimate, and $T_{\rm dust} = T_{\rm kin} = T_{\rm exc}$ under perfect
LTE, owing to the efficient dust and gas coupling \citep{carilli_2013,
dacunha_2013}. 
Here we identify a mild correlation between $T_{\rm exc}$ and $T_{\rm
  dust}$ only for the galaxy-integrated emission
  from local objetcs with secure \ci\ line detections
($\rho=0.47$, $0.62$, and $0.63$ Kendall, Spearman, and Pearson's
correlation coefficients, respectively), in agreement with previous
results, holding down to sub-galactic scales
\citep{jiao_2017, jiao_2019}. However, applying a
generalized Kendall's tau correlation coefficient to include the lower
limits on $T_{\rm exc}$ with the task \textit{bhkmethod} in IRAF
\citep{feigelson_1985}, we find significant probabilities that $T_{\rm
  exc}$ and $T_{\rm dust}$ are not correlated ($p=0.3071$). Similarly,
at this stage there are no hints of a strong correlation
between $T_{\rm  exc}$ and $T_{\rm dust}$ for the high-redshift
galaxies taken alone, nor for the
compilation as a whole, with similar probabilities from the
generalized Kendall's tau test or even considering detections
only. However, as clear from Figure
  \ref{fig:ciobsratios}, this result might stem from the relatively
  sparse high-redshift sample and its low number statistics. Notice
  that in the vast majority of the cases, we find $T_{\rm exc}<T_{\rm  dust}$.\\ 

On the contrary, a mild correlation is present when comparing
\lprimecoseven/\lprimecoone\ and \lprimecitwo/\lprimecione\
($p=0.0089$ probability that the two ratios do not  correlate from the generalized
Kendall's tau test; $\rho=0.39$, $0.52$, and $0.57$ Kendall, Pearson,
and Spearman's correlation
coefficients, respectively, considering only the detections, Figure
\ref{fig:ciobsratios}). When not directly
measured, for a handful of SMGs we
estimated \lprimecoone\ by converting \lprimecothree\ luminosities
following \cite{bothwell_2013}. No usable low-$J$ CO transitions are
available for our sample of main sequence galaxies at $z\sim1$ with
coverage of both \ci\ lines, excluded the aforementioned
\#35349. The
observed correlation suggests that the CO and \ci\ excitation
conditions are related and, by extension, that the temperature traced
by \ci\ increases for steeper CO SLEDs.

\section{Discussion}
\label{sec:discussion}

\subsection{A view on the ISM: modeling of the photodissociation regions}
\label{sec:modeling}
We now attempt to provide further insight into the physical conditions
of the ISM in galaxies by modeling the \ci, \co, and IR emission
following the classical recipes for photodissociation regions (PDRs).  
Here we adopt the one-dimensional models by
\cite{kaufman_1999} in the updated version released with the
\textsc{PhotoDissociation Region
  Toolbox}\footnote{\texttt{http://dustem.astro.umd.edu/pdrt}}
\citep[PDRT;][]{kaufman_2006, pound_2008}. The models solve
simultaneously for the chemistry, thermal balance and radiative
transfer, assuming metal, dust, and polycyclic aromatic hydrocarbons
(PAHs) abundances, and a gas microturbulent velocity dispersion. For
each combination of properties, a model is described in terms of the (number)
density of H nuclei $n \, [\mathrm{cm}^{-3}]$ and the intensity of the
incident far-ultraviolet radiation (FUV, $6 \, \mathrm{eV} < h \nu < 13.6 \,
\mathrm{eV}$) $G_{0}$ in units of the local Galactic interstellar
field \citep[$G_{0} = 1.6\times10^{-3}$ \esc,][]{habing_1968}. The
original models cover the $10 \leq n \leq 10^{7}$ cm$^{-3}$ and $10^{-0.5} \leq
G_0 \leq 10^{6.5}$ ranges in step of 0.25 dex. However, we rebinned
the templates to a $5\times$ finer grid before fitting the data: this did not affect
the final best-fit estimate of the density and intensity of the
radiation field, but it allowed us to assess the statistical
uncertainties of the fit. We computed the latter by applying the
$\chi^2$ criterion by \cite{avni_1976}, fixing $\Delta \chi^{2} <
2.71$ corresponding to a 90\% confidence interval. We also adopted a
purely numerical approach by bootstrapping 1000 times the observed
flux ratios within their errors and using the 68\%, 90\%, and 95\%
inter-percentile ranges as the corresponding confidence intervals. 
The best-fit model results from the $\chi^2$ minimization of the line and
continuum emission ratios. 

To the standard ratios available from PDRT, we added the \ci/IR
ratios. The latter mainly depend on $G_0$ through the IR emission due
to the dust clouds absorbing the UV incident emission and reprocessing
it at longer wavelength. On the other hand, in standard one-dimensional
PDR models, \ci\ arises from the
C$^+$/C/CO transition layer, which can be pushed deeper into the cloud
when the FUV field increases, but remaining substantially unchanged,
so that the column density of C does not depend on $G_0$
\citep[e.g.][]{tielens_1985, kaufman_1999, gerin_2000}.
We computed the IR intensity map as
$2 \times 1.3\times10^{-4} G_0$ erg cm$^{-2}$ s$^{-1}$ sr$^{-1}$,
including the contribution to the
global dust heating from of photons outside the FUV regime and
considering the finite slab
geometry \citep{kaufman_1999}. An extra factor of $2\times$ should be
included when considering the case of multiple clouds filling the
beam -- as for our unresolved measurements --, being illuminated from
every side. In this case, the optically thin IR emission from both the
near and far side of clouds would be visible
\citep{kaufman_1999}. However, a similar factor applies to
the optically thin \ci\ emission, canceling out this effect. 
We adopted the total IR luminosity \lir(8-1000 $\mu$m) due to
star-formation (i.e., removing the AGN contribution) from the
SED modeling as the estimate for IR. The only exceptions are
high-redshift QSOs, where the AGN emission dominates the far-IR
SED and we could not distinguish the contribution from star
formation. We therefore used the total \lir\ and we highlighted their
location in the relevant plots. As noted above, systematic deviations in the \ci/IR
ratios are largely due to this effect (V18).
Notice also that the \cite{draine_2007} suite of templates accounts for the independent
contributions to the total emission from the diffuse ISM and the PDRs,
but the available data does not allow us to discriminate between these
two components. We therefore used the combined, total IR
emission for the modeling.\\ 

\begin{figure*}
  \centering
  \includegraphics[width=0.48\textwidth]{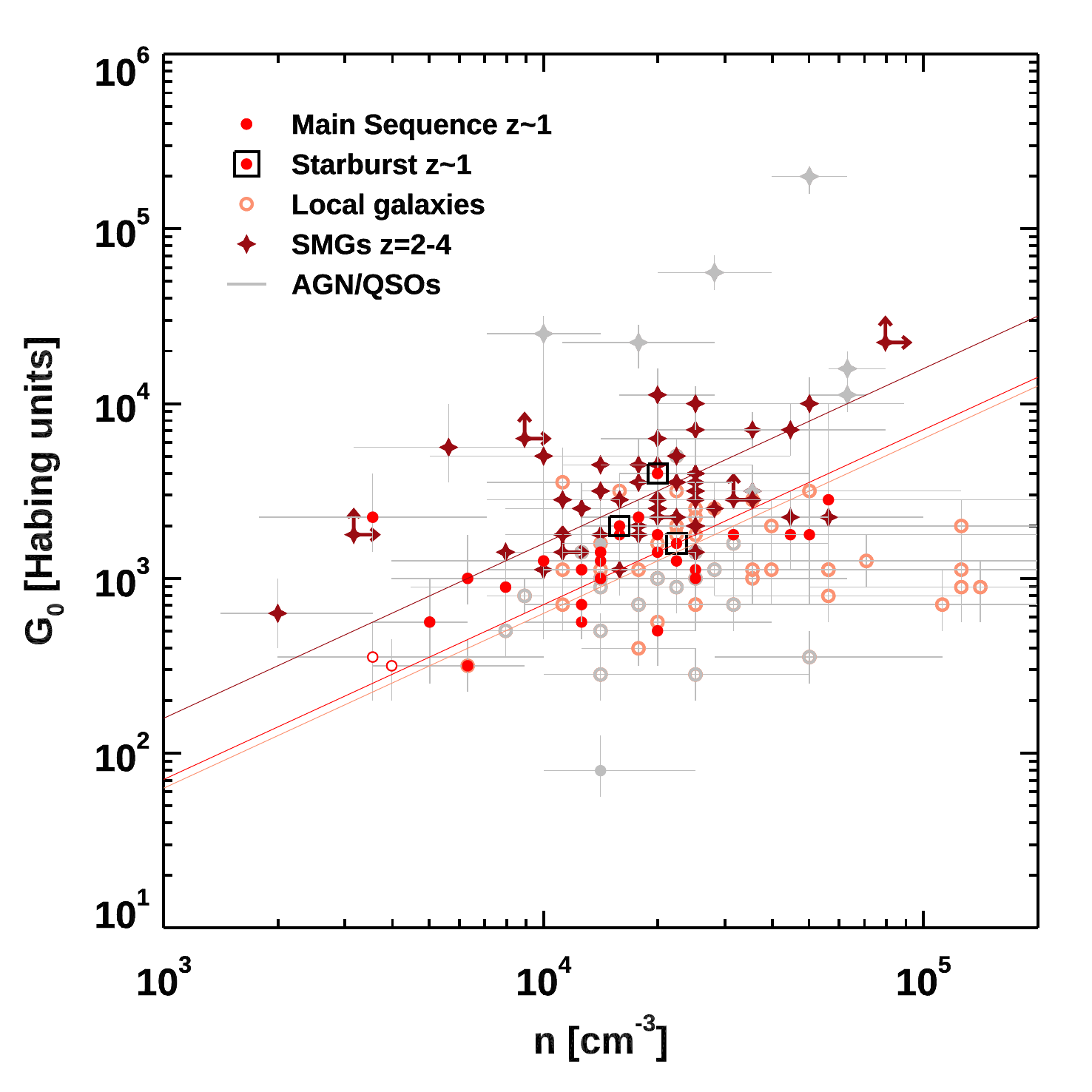}
  \includegraphics[width=0.48\textwidth]{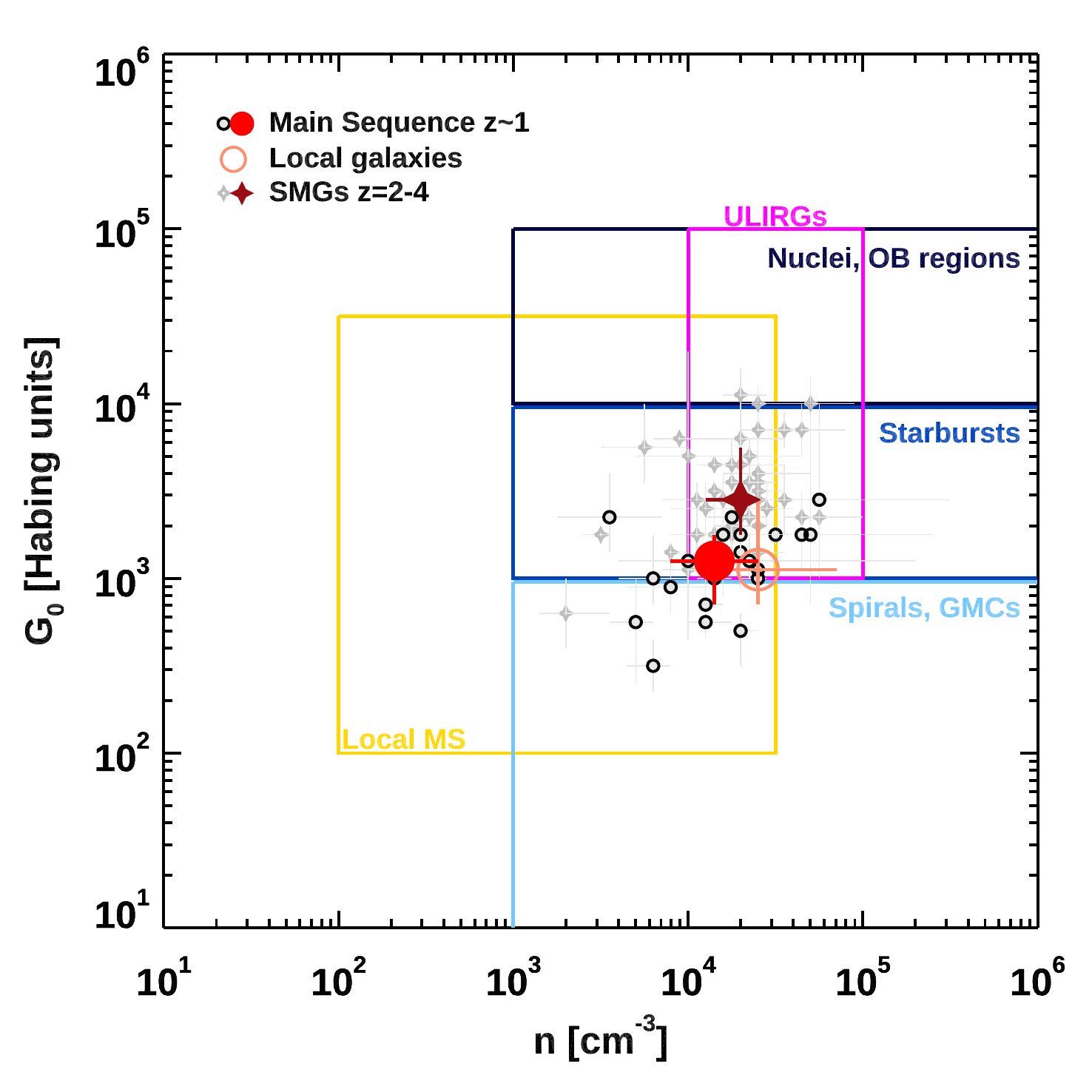}
  \caption{\textbf{Modeling of the photodissociation regions.}
   \textit{Left}: Best-fit gas density $n\,\mathrm{[cm^{-3}]}$ and intensity of the UV radiation field
    $U_{\rm UV}\, [\rm{Habing \; units, G_0}]$ from the PDR models
    by \citet{kaufman_1999} on the observed \cione, \cofour, and the
    IR estimates. Colors and symbols are as in Figure
    \ref{fig:obsratios}.  The dark, red, and light solid lines mark
    the loci of constant $G_{0}/n$ equal to the medians of the SMG,
    main sequence, and local galaxies, respectively. \textit{Right}:
    Median locations of the main sequence galaxies (red solid circle),
    local starbursts sample (open orange circle), and SMGs at $z\sim2-4$
    (dark red solid star) in the same plane as in the left panel. The
    error bars on the median represent the 16\% - 84\%
    inter-percentile range. For
    reference, we show the location of the individual main sequence
    galaxies (black open circles) and the SMGs (gray stars). The blue
    solid contours enclose the regions occupied by local galactic
    nuclei/OB regions, Starbursts, and Spirals/GMCs from
    \citet{stacey_1991}. The yellow square indicate the position
  of local main sequence galaxies from \citet{malhotra_2001}. The
  purple box shows the location of ULIRGs from \citet{davies_2003}.}
  \label{fig:pdrmodeling}
\end{figure*}

While $G_0$ is constrained by \ci/IR, a \ci/mid- or high-$J_{\rm
  upper}$ CO ratio is an effective tracer of the gas density, 
being almost insensitive to $G_0$. Therefore, the
  combination of \ci/IR and \ci/mid- or high-$J_{\rm upper}$ CO allows for a full
  determination of the PDR properties. This is clear from the nearly
perpendicular tracks in Figure \ref{fig:obsratios}\footnote{The
  intensity ratio maps from the \cite{kaufman_1999} models are
  available in .fits format in the original PDRT
  website.}. Notice that our grid of models is different from the one
in \citet{alaghband-zadeh_2013} due to the diverse approaches to
map \lir\ into $G_0$ (Figure \ref{fig:ci_test} in Appendix).
These models were applied to the available combinations of observed
ratios. We show the $n, G_0$ values from the modeling of \lcione,
\lcofour, and \lir\ in Figure \ref{fig:pdrmodeling}. The
median location of the local LIRGs, main sequence galaxies, and
high-redshift SMGs are also shown in comparison with regions occupied
by local main sequence galaxies \citep{malhotra_2001}, spirals and giant molecular clouds,
starbursts, nuclei and OB regions \citep{stacey_1991}, and ULIRGs
\citep{davies_2003}. The conclusions about the observed line ratios are
naturally reflected on the similar $G_0/n$ ratios for main sequence
galaxies at $z\sim1$ and local LIRGs, both lower than for SMGs. The
trend is driven by the increasing $G_0$
($\propto($\lcione/\lir$)^{-1}$). The median location and the
distribution of the $n, G_0$ values for main sequence galaxies at
$z\sim1$ is also consistent with the approximate boundaries for local
similar samples and spiral and GMCs, areas completely devoided of
SMGs.\\

Therefore, interpreting the gas and dust emission according to
photodissociation region modeling further suggests similar ISM
conditions in main sequence galaxies and local LIRGs, both less
extreme that in high-redshift SMGs.  

\subsubsection{Caveats and shortcomings of PDR modeling}
The modeling presented above allows only for a
simple interpretation of basic combinations of line ratios, especially
considering  the lack of spatial resolution. However, the trends
identified in the previous Section should be treated as an
order of magnitude indication, since the modeling suffers from multiple
limitations, as previously noted by several authors who attempted a
similar analysis on part of the samples collected here
\citep[Appendix \ref{sec:appendix},][and references
therein]{alaghband-zadeh_2013, bothwell_2017, canameras_2018}. Here we
report further evidence that the view on the ISM offered by
simple one-dimensional modeling is likely not sufficient to capture
the physical complexity of this medium. As shown in Figure 
\ref{fig:obsratios}, galaxies in the \cione/\cofour/IR and
\citwo/\coseven/IR planes are mapped into different regions of the
$n, G_0$ parameter space, the former line combination returning lower
densities and intensities of the FUV radiation fields than the
latter. This likely reflects gas in different phases, warmer and
denser in the \citwo/\coseven/IR diagram than in
\cione/\cofour/IR. The adopted models struggle to reproduce the
densest molecular gas, given the assumption of a moderate constant attenuation $A_{\rm
  V}=1.8$ mag. We quantified the effect of modeling \cione\ or \citwo\
keeping \co+IR fixed for the galaxies with both \ci\ lines
available. Figure \ref{fig:ci_test} in Appendix shows that using \citwo\ results
in $\sim0.5$ and $\sim0.3$~dex larger $n$ and $G_0$ than when adopting
\cione\, respectively, explaining most of the shift observed in Figure
\ref{fig:obsratios}. We similarly tested the impact of using \ciplus\
in lieu of \lir\ for a sub-sample of the SPT-SMGs at $z\sim4$. Both
these quantities are primarily dependent on $G_0$, so $n$ is not
affected by this choice. Figure \ref{fig:ci_test} shows that, despite
the large scatter, modeling \ciplus\ or \lir\ gives consistent
results. These tests show that the larger spread of $n, G_0$ reported when
fitting simultaneously all the available luminosity ratios
\citep{danielson_2011, alaghband-zadeh_2013} likely
results from the inability of a single PDR model to capture a
multi-component ISM.
\begin{figure*}
  \centering
  \includegraphics[width=\textwidth]{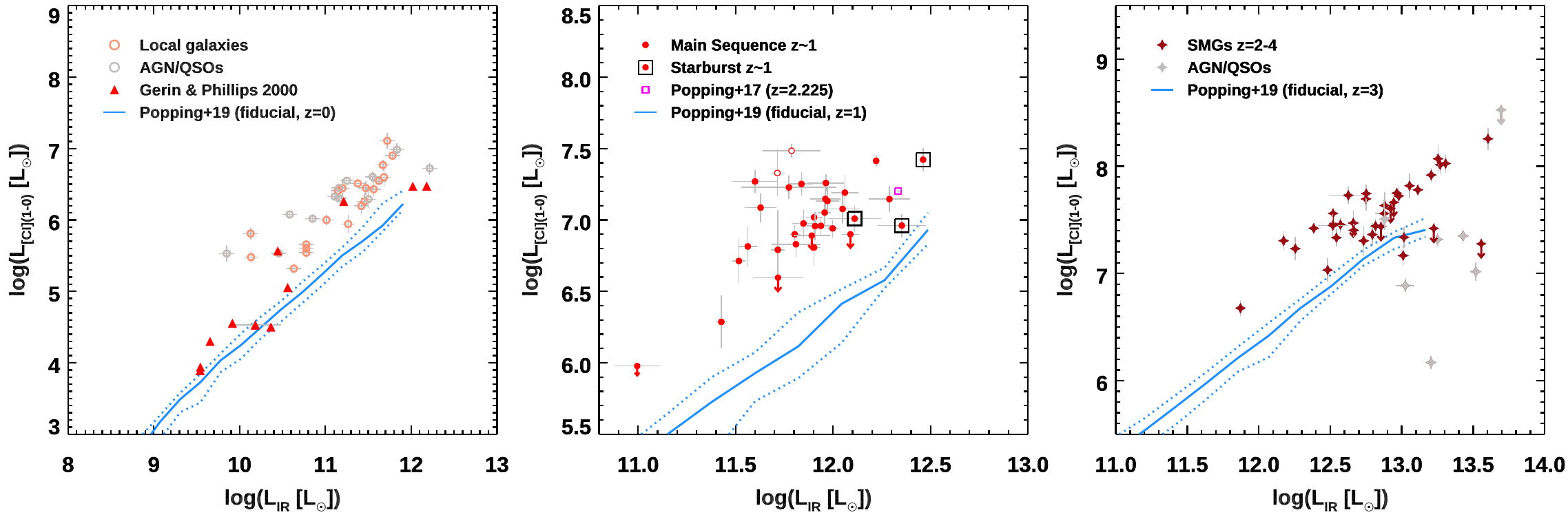}
  \caption{\textbf{Semi-analytical modeling of \cione\ emission in
      galaxies.} \textit{Left:} Empty orange and gray circles mark the
    local sample of starburst galaxies without and with AGN signatures
    The red filled triangles show the galaxies from
    \citet{gerin_2000}. \textit{Center}: The red filled circles show
    the main sequence galaxies at $z\sim1$ (red empty circles: sources
    with caveats from \citealt{bourne_2019}). The empty black squares
    indicate the starbursts at
    $z\sim1.2$. To highlight the different redshift, the empty
    purple square marks the main sequence object at
    $z=2.225$ from \citet{popping_2017}. \textit{Right:} The dark red
    solid stars show $z\sim2-4$
    SMGs 
The gray solid stars indicate
    QSO/AGN-dominated galaxies. Arrows mark $3\sigma$ upper limits on
    \ci. In every panel, the blue solid and dotted lines indicate the fiducial
    model by \citet{popping_2019} and its 16\% and 84\%
    percentiles at the average redshift of the samples. 
}
  \label{fig:sammodeling_ci10}
\end{figure*}
\begin{figure*}
  \centering
  \includegraphics[width=\textwidth]{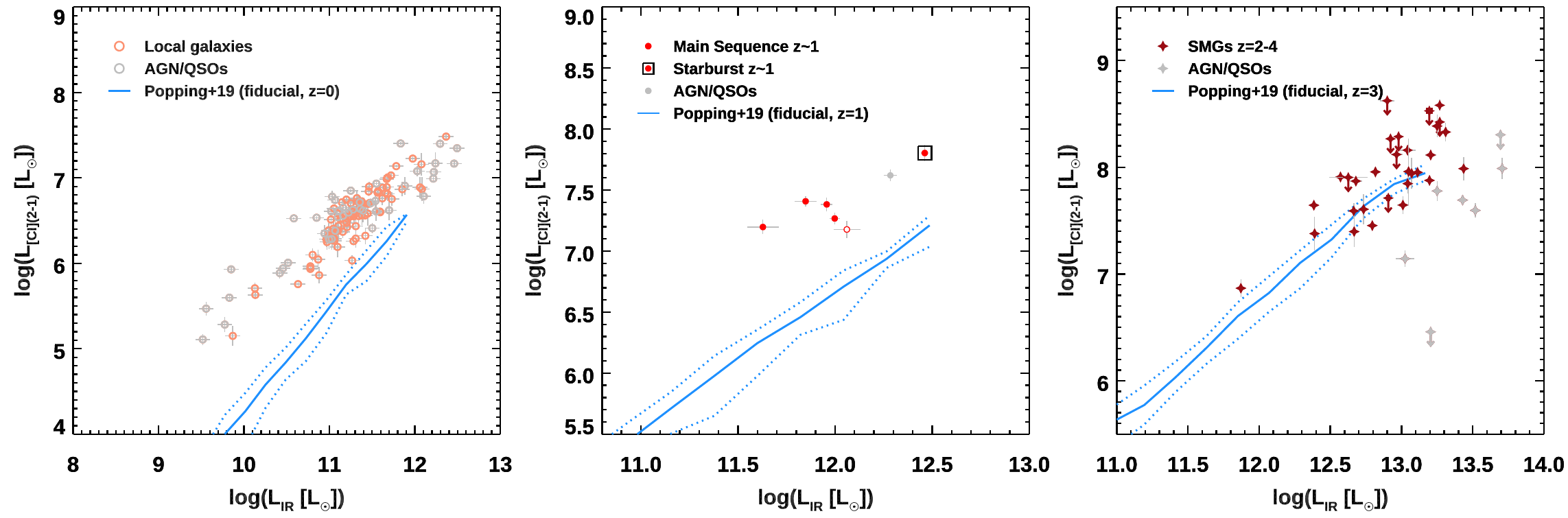}
  \caption{\textbf{Semi-analytical modeling of \citwo\ emission in
      galaxies.} Symbols and colors indicate galaxies as in Figure
    \ref{fig:sammodeling_ci10}, when a \citwo\ estimate is
    available. We did not include any upper limits for the local
    sample in order to avoid crowding the panel.}
  \label{fig:sammodeling_ci21}
\end{figure*}
\begin{figure*}
  \centering
  \includegraphics[width=\textwidth]{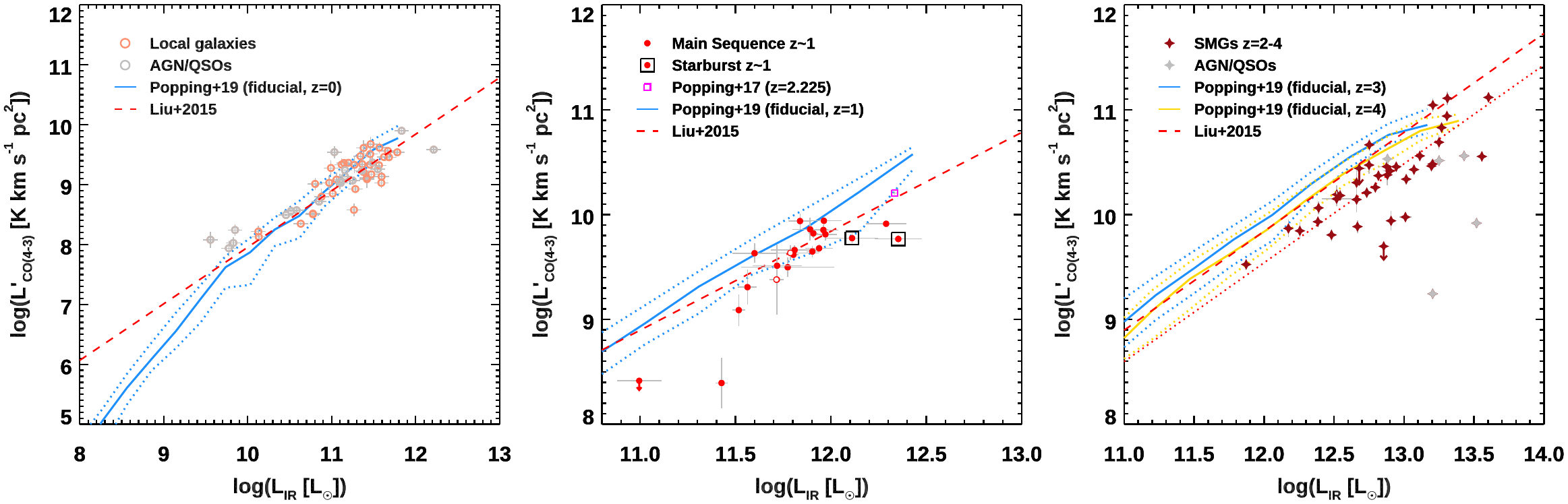}
  \caption{\textbf{Semi-analytical modeling of \cofour\ emission in
      galaxies.} Symbols and colors indicate galaxies as in Figure
    \ref{fig:sammodeling_ci10}, when a \cofour\ detection is
    available (no correction from other \co\ transition is shown
    here). In every panel, the dashed red lines indicate the
    \lir-\lprimecofour\ correlation from \citet{liu_2015}. In the
    right panel, the blue and gold lines indicate the
    fiducial models from \citet{popping_2019} and their 16\% and 84\%
    percentiles at $z=3$ and $z=4$, respectively. The red dotted line
    indicates the \citet{liu_2015} correlation scaled by $-0.3$~dex as
    reported in the original paper for starbursts/SMGs.}
  \label{fig:sammodeling_co43}
\end{figure*}

\subsection{Gas and dust temperatures}
\label{sec:temperatures}
In Section \ref{sec:ciratio} we reported the
  presence of a mild correlation
between the gas excitation temperature $T_{\rm exc}$ derived from the
\ci\ line ratio and the luminosity-weighted dust temperature $T_{\rm dust}$ (or the
mean intensity of the radiation field $\langle U \rangle$, Figure
\ref{fig:ciobsratios}) for local galaxies, but
  also the lack of indications that a similarly significant correlation is in place
for the high-redshift  objects, possibly because of the sparsity of
such sample. We
further found that, generally, $T_{\rm
  exc} \lesssim T_{\rm dust}$ in our compilation, with a rather
constant $T_{\rm exc}$ across the different populations and redshifts.
If confirmed,
this can be explained considering that, in first approximation, \ci\
arises from a thin transitioning layer between \co\ and \ciplus, being
insensitive to the ambient UV radiation field. Moreover, the fact that
$T_{\rm exc} \lesssim T_{\rm dust}$ is consistent with what is
reported for a subsample of high-redshift SMGs collected here \citep{nesvadba_2018}. This
might suggest that gas and dust are not in thermal equilibrium
\citep{canameras_2015, nesvadba_2018}. Alternatively,
considering a more realistic multi-phase ISM, $T_{\rm dust}$ might be
determined by a hot component dominating the far-IR emission
close to the peak and physically closer to the starbursting regions, while \ci\
and the cold dust extend further out, tracing the bulk of the mass of
the molecular gas \citep[V18,][]{nesvadba_2018}. 
Our findings agree
with recent findings by \citet{jiao_2019}, who 
retrieve a moderate $T_{\rm exc}$-$T_{\rm dust}$ correlation in
resolved maps of local galaxies (Section \ref{sec:ciratio}).
We do find $T_{\rm exc}\lesssim T_{\rm dust}$ for the high-redshift
sample as for the local resolved objects \citep{jiao_2019_erratum}
and compatibly with the results by
  \cite{bothwell_2017} based on a line
modeling approach.
Here we note that, while $T_{\rm exc}$ is derived in a
consistent way across different papers, $T_{\rm dust}$ is highly
susceptible of strong variations due to the available photometry and
the adopted parametrization of the IR SED. In particular, the
lack of coverage of the peak of the emission 
strongly affects the $T_{\rm dust}$ estimate while classical
single-temperature modified black body curves cannot reproduce
the observed mid-IR emission, suggesting the
existence of multi-component dust along with the ISM
\citep{draine_2007, galliano_2011, magdis_2012, casey_2014,
  schreiber_2018_dust, liang_2019}. \citet{jiao_2019} estimate $T_{\rm
  dust}$ by assuming a gray body with $\beta= 2$ modeling the rest-frame
$70/160$~$\mu$m ratio, a color that we cannot directly measure at high
redshift. On the other hand, we fit the whole available far-IR SED with a
single-temperature MBB leaving $\beta$ free to vary and assuming an
optically thin emission, a common choice allowing for a direct
comparison with data in the literature. This latter condition might
have to be reconsidered, especially
for strongly starbursting objects and SMGs. As shown by Cortzen et
al. (submitted), removing this constraint in the SED modeling allows them
to derive $T_{\rm dust}=52\pm5$~K for GN20, the strongest outlier in
Figure \ref{fig:ciobsratios}. This is more consistent with $T_{\rm
  exc}=48^{+15}_{-9}$~K from \ci\ than in the optically thin case
shown in Figure \ref{fig:ciobsratios} ($T_{\rm
  dust}=33\pm2$~K). At the current stage, when removing the $T_{\rm exc}\gg
T_{\rm dust}$ outliers, the correlation between these two quantities
in the high-redshift sample remains
weak. The possible future extension of the
  results on the optical depth of GN20 to the
general population of strongly starbursting systems at high redshift
(including the role of the cosmic microwave background \citealt{dacunha_2013, zhang_2016}),
might change this conclusion \citep[e.g.,][]{jin_2019}. Finally, we
note that we consistently modeled the dust emission for all galaxies
at different redshifts. If a significant evolution with time were
present, e.g., due to a metallicity change, this could affect the
current results. However, given the large stellar masses and SFRs of
the objects collected in this work, metallicity is unlikely to play a
major role.

\subsection{Neutral atomic carbon in a cosmological galaxy evolution context}
\label{sec:popping}
The access to a large statistical sample of galaxies covering wide
ranges of redshifts and physical conditions allows us to start exploring the role 
played by neutral atomic carbon in a broader cosmological context of galaxy
evolution, an operation so far accessible for a few \co\
transitions, dust, and increasingly for the bright \ciplus\ emission thanks to
similarly numerous samples coming online \citep[e.g.,][to mention a
few recent efforts]{carilli_2013,
  tacconi_2018, zanella_2018, liu_2019}. Here we compared the \ci\ and \co\ line 
luminosities from our compilation with the semi-analytical model
from \citet{popping_2019}. Briefly, the authors adopted the latest version of
the `Santa Cruz' galaxy formation model \citep{somerville_1999,
  somerville_2001, popping_2014, somerville_2015} as the input to
shape the emission of sub-mm \co,
\ciplus, and \ci\ lines \citep{krumholz_2014, narayanan_2017}. They
applied various subgrid recipes to describe the dense and diffuse gas
distribution, the density profile within molecular clouds, the
clumping of the medium, the UV and cosmic ray fluxes regulating the
ionization and chemistry of the clouds. By combining chemical
equilibrium networks and radiative transfer models with sub-grid
models, \citet{popping_2019} finally obtained different sets of \co, \ciplus,
and \ci\ luminosities emerging from galaxies and readily comparable
with observations. We refer the reader to the original paper for
further details.\\

Figures \ref{fig:sammodeling_ci10}, \ref{fig:sammodeling_ci21}, and
\ref{fig:sammodeling_co43} show the \cione\, \citwo\ and \cofour\
luminosities of our samples of galaxies compared with the predictions
from the fiducial model by \citet{popping_2019}. We show the samples
divided in redshifts bins and galaxy type (local LIRGs, main sequence
galaxies at $z\sim1$, and SMGs at $z=2-4$, Section
\ref{sec:data}). As in previous plots, we also show the position of
galaxies with contamination from AGN/QSOs. For reference, in Figure
\ref{fig:sammodeling_ci10} we display the local galaxies from
\citet{gerin_2000} originally reported in \citet{popping_2019}. We
further remind the reader that we corrected both the measurements of
the dust and line emission of the local galaxies for the aperture
correction \citep[V18,][]{liu_2015}. Such correction is identical for
\lir\ and the line luminosity, therefore moving the galaxies
diagonally in the panels. Moreover, we converted the tracks originally
expressed as a function of SFR into \lir\ by applying the
\citet{kennicutt_1998} conversion for a \citet{chabrier_2003} initial
mass function. \citet{popping_2019} originally
  carried out the comparison with the observations by converting the
  \lir\ to SFR following the relation in \cite{murphy_2011}. Using the
  latter results in 0.16 dex lower \lir\ for a fixed SFR than adopting
  \citet{kennicutt_1998}, not changing the substance of our results.
In every panel of Figures
\ref{fig:sammodeling_ci10} and \ref{fig:sammodeling_ci21} the observed
\ci\ luminosities appear brighter than the predictions of the fiducial
model at fixed \lir. The observations seem to follow a steeper
(shallower) trend than the model in the \lir-\lcione\ (\lcitwo) plane
at $z\sim0$, while the slope is similar at higher redshifts.
Notice that this tension would increase if a
  substantial fraction of the total SFR is unobscured, reducing \lir\
  and moving the points towards the left.  We register
the minimal discrepancy between \ci\ observations and models for
the high-redshift SMGs in the \lir-\lcitwo\ panel (Figure
\ref{fig:sammodeling_ci21}). 
On the contrary, the model successfully
reproduces the \cofour\ emission from local and main sequence galaxies
at $z\sim1$ (Figure \ref{fig:sammodeling_co43}). For reference, 
we also show the best fit \lir-\lprimecofour\ relation from
\citet{liu_2015}. This is not surprising, considering the 
performances of the fiducial model for low- and mid-$J$ \co\
transitions already reported by \citet{popping_2019} for similar
datasets. We draw similar conclusions for \cofive. On the contrary,
the model overpredicts the \lprimecofour\ luminosity of high-redshift
SMGs, which we find consistent with the -0.3~dex offset to the local
relation \citep{liu_2015, yang_2017}.\\

The offset between the \ci\ observations and the fiducial model, not
corresponding to a displacement of the \co\ and \ciplus\
measurements \citep[and references therein]{popping_2019}, suggests
that the emission of neutral atomic carbon is not fully captured by
the current recipes. We note that the
  sSFRs predicted by the Santa Cruz model do match the observed values for the massive star-forming
  population at $z=0$, but the normalization of the main sequence falls below the
  empirical estimates at $z>1$. This is a well known issue, as shown in Figure 11
  from \cite{somerville_2015}. Our sample of main sequence galaxies at
  $z\sim1$ makes no exception: the observed median
  $\mathrm{log(sSFR/yr^{-1})}=-9$ is larger than what is predicted by the
  model. However, the results of the comparison do not change even
  when limiting the model predictions to the star-forming population
  matching the observed sSFR threshold.
  On the other hand, the depletion timescales
  or SFEs are reasonably well described by the Santa
Cruz model
  \citep{somerville_2015, popping_2019_aspecs}.
 Systematic differences in SFEs are unlikely to drive the
  discrepancy with the observations in Figures
  \ref{fig:sammodeling_ci10}-\ref{fig:sammodeling_co43}, since both
  \ci\ and \co\  should have been similarly impacted. While
    further work on the model is necessary to 
    remove the systematics on the sSFR, the relative comparison of
    \co\ and \ci\ emission as a function of \lir\ still holds.
  Therefore, the problem likely arises from the modeling of the
  emission itself.\\

Interestingly, the \ci\ emission appears to be
largely unaffected by several parameters, including the
the density of the diffuse atomic ISM, the choice of rescaling the
strength of the UV- and cosmic
ray fields to the local or the global SFR, the slope of the molecular
clouds distribution, the clumping of the ISM, and the radial density
profile within the clouds at a fixed external pressure. On the
contrary, the choice of the density profile within the molecular clouds 
might increase the \lci\ at fixed SFR (Figure 9 in
\citealt{popping_2019}), also modifying the slope
of the relation, as the observations suggest (Figures
\ref{fig:sammodeling_ci10} and \ref{fig:sammodeling_ci21}).
Nevertheless, modifying only this parameter would generate tensions
with the CO and \ciplus\ observations that are currently indiscernible. We
underline the fact that the model is meant to reproduce the bulk of
the galaxy population. Therefore, strongly deviating outliers, such as
starbursts and SMGs in the standard definitions
and shown here for the sake of completeness, would likely require
a specific treatment. \\ 

Further developments of these and alternative models appear necessary in
order to reproduce the observations. The compilation we publicly
release here will serve as a useful tool for calibration.\\

\section{Conclusions}
\label{sec:conclusions}
We presented new observations of \citwo\ and \coseven\ in a sample of $7$
main sequence galaxies at $z\sim1$, along with a compilation of
$>200$ objects with detected \cione\ and/or \citwo, and one or multiple low to high-$J$ CO
transitions. This compilation spans the $z=0-4$ redshift interval and a wide range of
ambient conditions. We leveraged such a large sample to derive the
following results. 
\begin{itemize}
\item The \lcione/\lir\ and \lcitwo/\lir\ ratios in local LIRGs and
  main sequence galaxies at $z\sim1$ are consistent, but
  systematically and significantly higher than the values measured in
  SMGs at $z=2-4$. We draw a similar conclusion for the
  \lcitwo/\lcoseven\ ratio, while \lcione/\lcofour\ appears more
  consistent among the different redshifts and populations, resembling
  the case of \cotwo\ previously reported in V18.
\item The previous point can be interpreted considering that \ci,
  mid-/high-$J$ CO ($J_{\rm upper}=3-7$), and \lir\ trace the low
  density, high density molecular gas,
  and the SFR, respectively. The observed trends would thus suggest
increased dense molecular gas fractions and higher SFEs in high-redshift SMGs than
main sequence galaxies and local LIRGs, in agreement with CO-based
studies. This is further consistent with the (integrated) Schmidt-Kennicutt
relation and serves as supporting evidence of the potential of
\ci\ as an alternative molecular gas tracer.
\item The observed \lprimecitwo/\lprimecione\ ratios of the local,
  main sequence at $z\sim1$ and high-redshift SMG samples are fully
  consistent. This suggests a similar gas temperature traced by \ci\
  across redshift and galaxy type. 
\item Moreover, the \lprimecitwo/\lprimecione\ ratio ($\propto T_{\rm
    exc}$) mildly correlates with \lprimecoseven/\lprimecoone, a proxy
  for the shape of the CO SLED. More extreme excitation for
  \ci\ corresponds to similar conditions of CO, supporting the
  coexistence of these molecular gas tracers.
\item We find $T_{\rm exc}\lesssim T_{\rm dust}$ on global scales for
  the galaxies in our compilation and  a mild correlation between these
  two quantities for local galaxies, but no apparent strong connection
  in the high-redshift sample. 
  This might be due to the presence of
  multi-phased ISM and dust, with \ci\ tracing a cold and extended
  component and the (luminosity-weighted) $T_{\rm dust}$ from the SED modeling being mainly
  driven by hot dust, not accounting for the bulk of the dust
  mass in the cold phase. However, this result critically depends on the modeling
  and parametrization of the far-IR SED and on 
  relatively small number statistics of galaxies with both \ci\ lines
  available, especially at high redshift.  
\item We fitted the observed line ratios using classical one-dimensional
  photodissociation region models \citep{kaufman_1999}, retrieving
  similar $n, G_0$ for local LIRGs and main sequence galaxies at
  $z\sim1$, but larger values for high-redshift SMGs, consistent with
  the conclusions reported above. 
\item While valuable for a simple interpretation and relative
  comparisons among spatially unresolved observations, classical 1D
  PDR modeling is hampered by several shortcomings. Here we explored
  the impact of using different combinations of line ratios, finding
  the necessity of a more comprehensive approach in order to reproduce
  a likely complex and multi-phased ISM at any redshifts. This adds to
  other well known issues previously reported.
\item We compared our compilation with available semi-analytical
  models \citep{popping_2019} in order to place \ci\ in a context of
  cosmological galaxy evolution. While we confirm a good agreement
  with fiducial model for the \co\ emission, we find systematically
  larger \ci\ luminosities at fixed \lir\ for any sample at any
  redshifts than the theoretical predictions. This suggests the
  necessity of a revision
  of some of the subgrid recipes assumed in the model.
\item We release the data compilation in an electronic format to the
  community as a benchmark for future testing and comparison.
\end{itemize}

\section*{Acknowledgements}
We acknowledge the constructive comments from the anonymous referee,
which helped improving this work.
We thank Mark Sargent and Anna Cibinel for providing the initial
catalogs for the selection of main sequence galaxies. 
We thank Nathan Bourne,
Fabian Walter and Axel Wei{\ss} for providing the photometry and line
emission for their samples. FV thanks Qian Jiao for clarifying details
of her work and sharing her data; Diane Cormier for discussions about the PDR
modeling. FV and GEM acknowledge the Villum
Fonden research grant 13160 ``Gas to stars, stars to
dust: tracing star formation across cosmic time'' and the Cosmic Dawn Center
of Excellence funded by the Danish National Research Foundation under then grant No. 140.
FV acknowledges support from the Carlsberg Foundation research grant CF18-0388 ``Galaxies: Rise And Death''.
GEM acknowledges support from the European Research Council (ERC)
Consolidator Grant funding scheme (project ConTExt, grant number
648179). SJ acknowledges financial support from the Spanish Ministry
of Science, Innovation and Universities (MICIU) under grant
AYA2017-84061-P, co-financed by FEDER (European Regional Development
Funds). DL acknowledges funding from the European Research Council
(ERC) under the European Union's Horizon 2020 research and innovation
programme (grant agreement No. 694343). YG's research is supported by
National Key Basic Research and Development Program of China (grant
No. 2017YFA0402704), 
National Natural Science Foundation of China (grant Nos. 11861131007,
11420101002), and Chinese Academy of Sciences Key
Research Program of Frontier Sciences (grant No. QYZDJSSW-SLH008).
In this work we made use of \textsc{STSDAS}. \textsc{STSDAS}
is a product of the Space Telescope Science Institute, which is
operated by AURA for NASA. 
This paper makes
use of the following ALMA data: ADS/JAO.ALMA, \#2018.1.00635.S, \#2016.1.01040.S,
 \#2016.1.00171.S, \#2015.1.00260.S. ALMA is a
partnership of ESO (representing its member states), NSF (USA) and
NINS (Japan), together with NRC (Canada), MOST and ASIAA (Taiwan), and
KASI (Republic of Korea), in cooperation with the Republic of
Chile. The Joint ALMA Observatory is operated by ESO, AUI/NRAO and
NAOJ. In this work we made use of the COSMOS master spectroscopic
catalog, available within the collaboration and kept updated by Mara Salvato.

%% Bibliography
\bibliography{../bib_ism_valentino} 
\bibliographystyle{aasjournal}

\appendix
\section{Spectral energy distribution of the high-redshift sample}
\label{sec:app:sed_literature}
\begin{figure*}
  \centering
  \includegraphics[width=\textwidth]{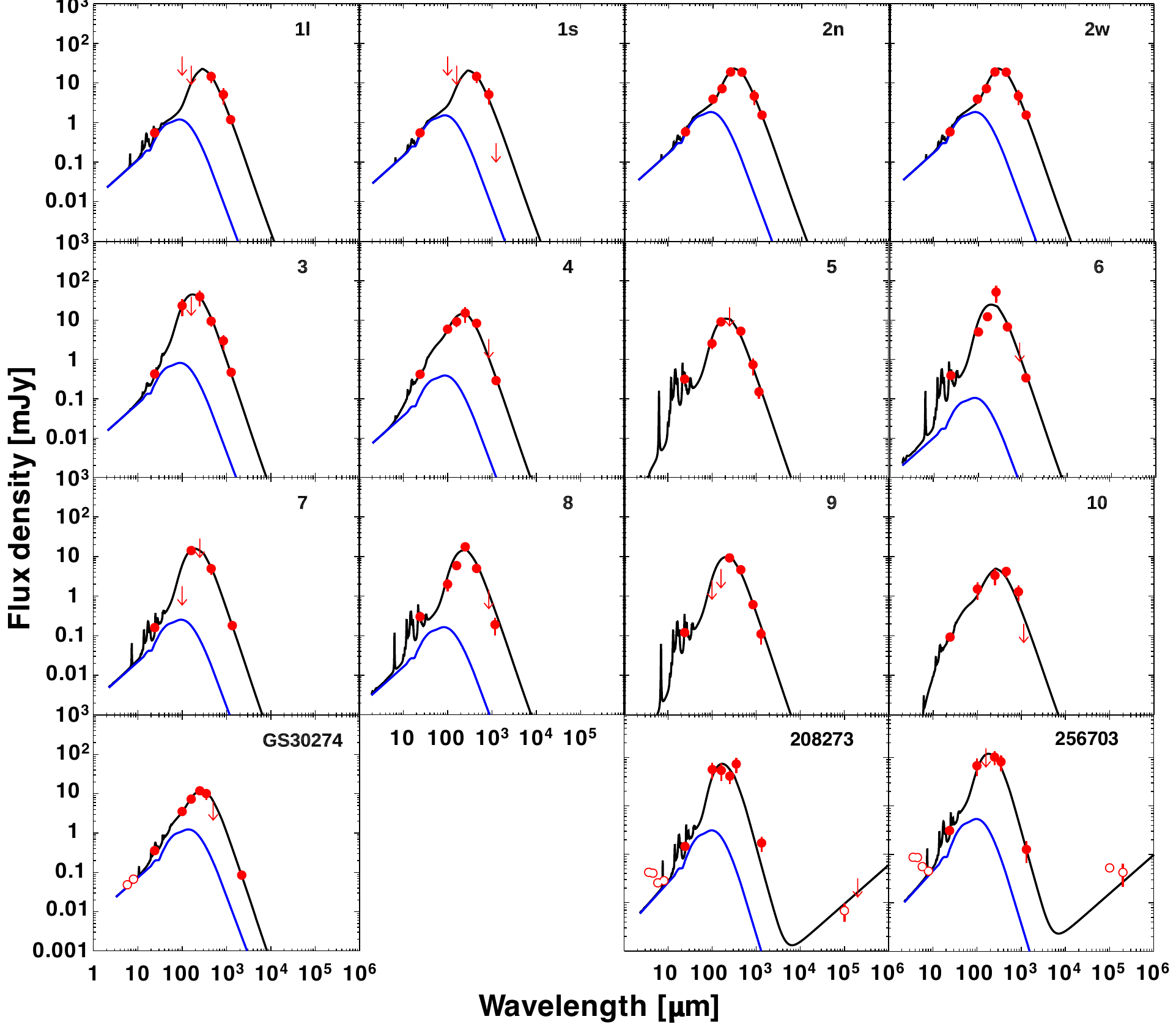}
  \caption{Far-infrared spectral energy distribution for the
    literature sample at high redshift. Red full circles mark the observed photometric
    points. Downpointing red arrows indicate $2\sigma$ upper
    limits. Open red symbols show photometry
    from \textit{Spitzer}/IRAC and radio when available, not used in
    the fit. The black line shows the best
    composite SED (\cite{draine_2007} and AGN templates from
    \cite{mullaney_2011}, the latter highlighted in blue).
    See Appendix \ref{sec:app:sed_literature} for details.}
  \label{fig:sed_literature}
\end{figure*}

 As part of the supplementary online material, we
  show the re-modeling of the SED we performed for all the
  high-redshift galaxies from the literature samples that we
  collected (Figure \ref{fig:sed_literature}). The SEDs of our
  own sample of main sequence galaxies are reported with an identical
  format in V18, excluding sources \#208273 and 256703 included in
  this release. We do not
  show the SMG from \cite{jin_2019}, being the modeling identical to
  the original work. In each panel, we show in red the photometric
  points we fitted and their uncertainties. Downpointing arrows
  indicate $2\sigma$ upper limits. Open red symbols mark photometry
  from \textit{Spitzer}/IRAC and radio when available, which we did
  not use to constrain the fit. The black line shows the best
  composite SED (\cite{draine_2007} and AGN templates from
  \cite{mullaney_2011}, the latter highlighted in blue).
\setcounter{figure}{6}    
\begin{figure*}
  \centering
  \includegraphics[height=\textheight]{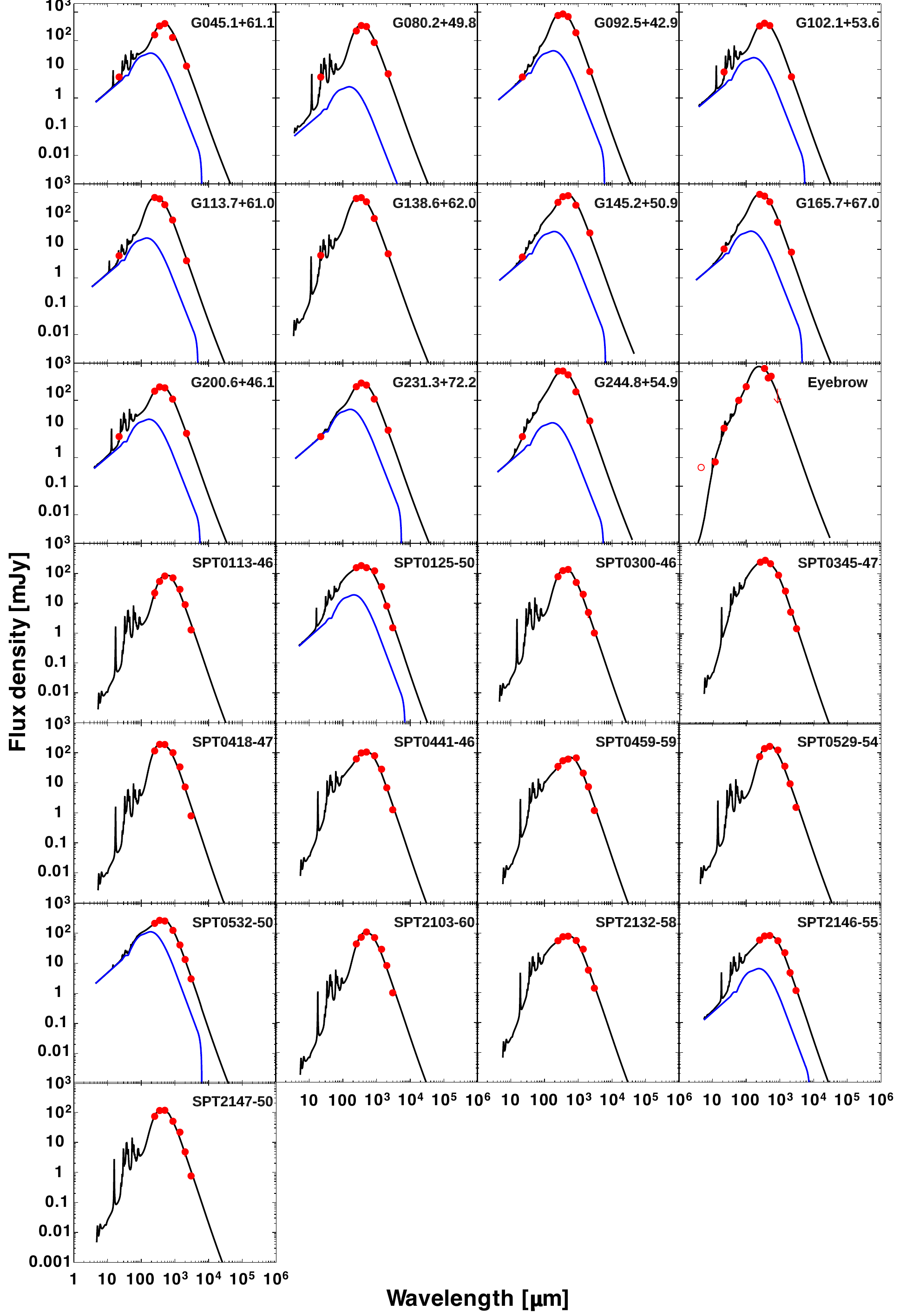}
  \caption{(continue)}
\end{figure*}
\setcounter{figure}{6}    
\begin{figure*}
  \centering
  \includegraphics[height=\textheight]{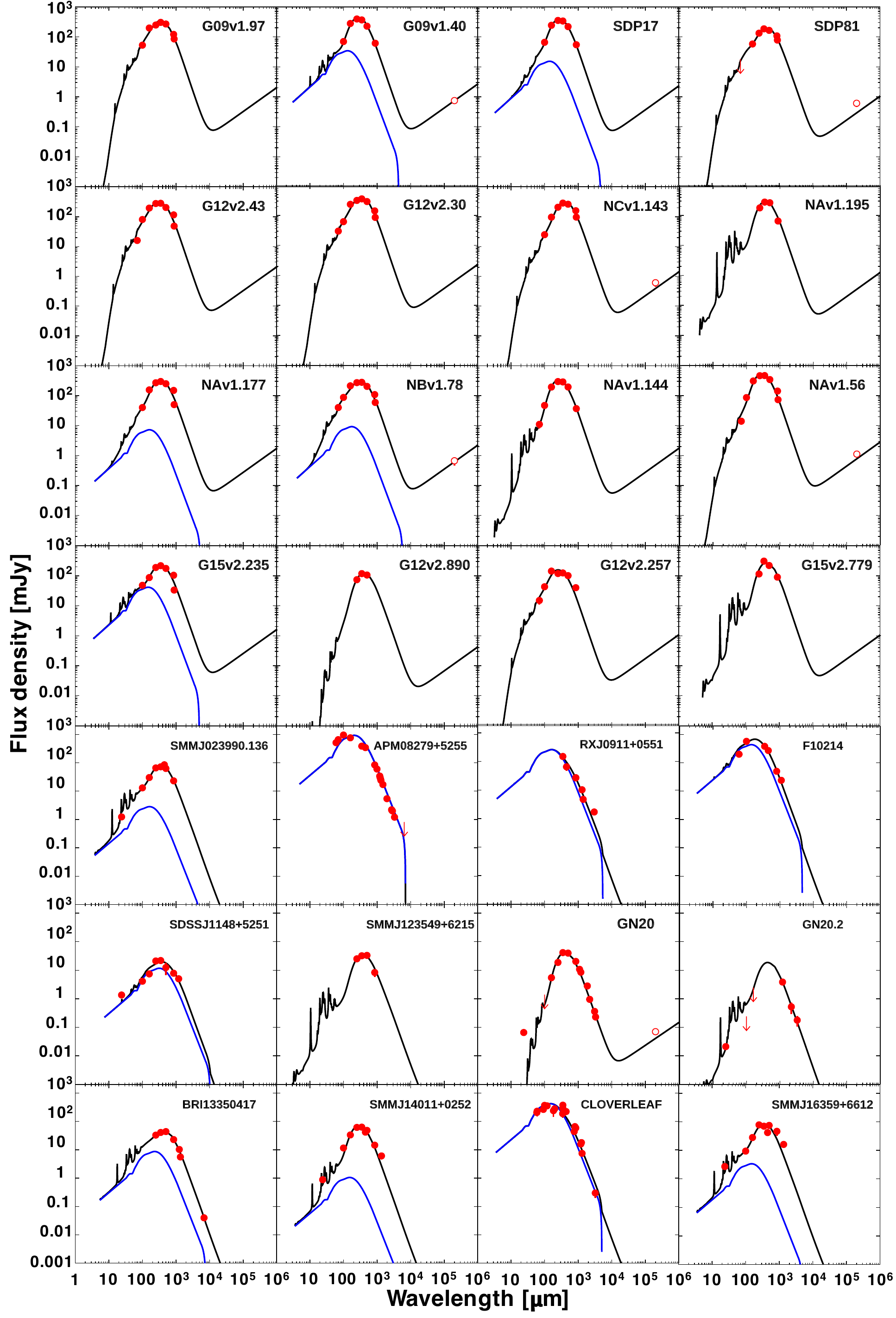}
  \caption{(continue)}
\end{figure*}
\setcounter{figure}{6}    
\begin{figure*}
  \centering
  \includegraphics[width=\textwidth]{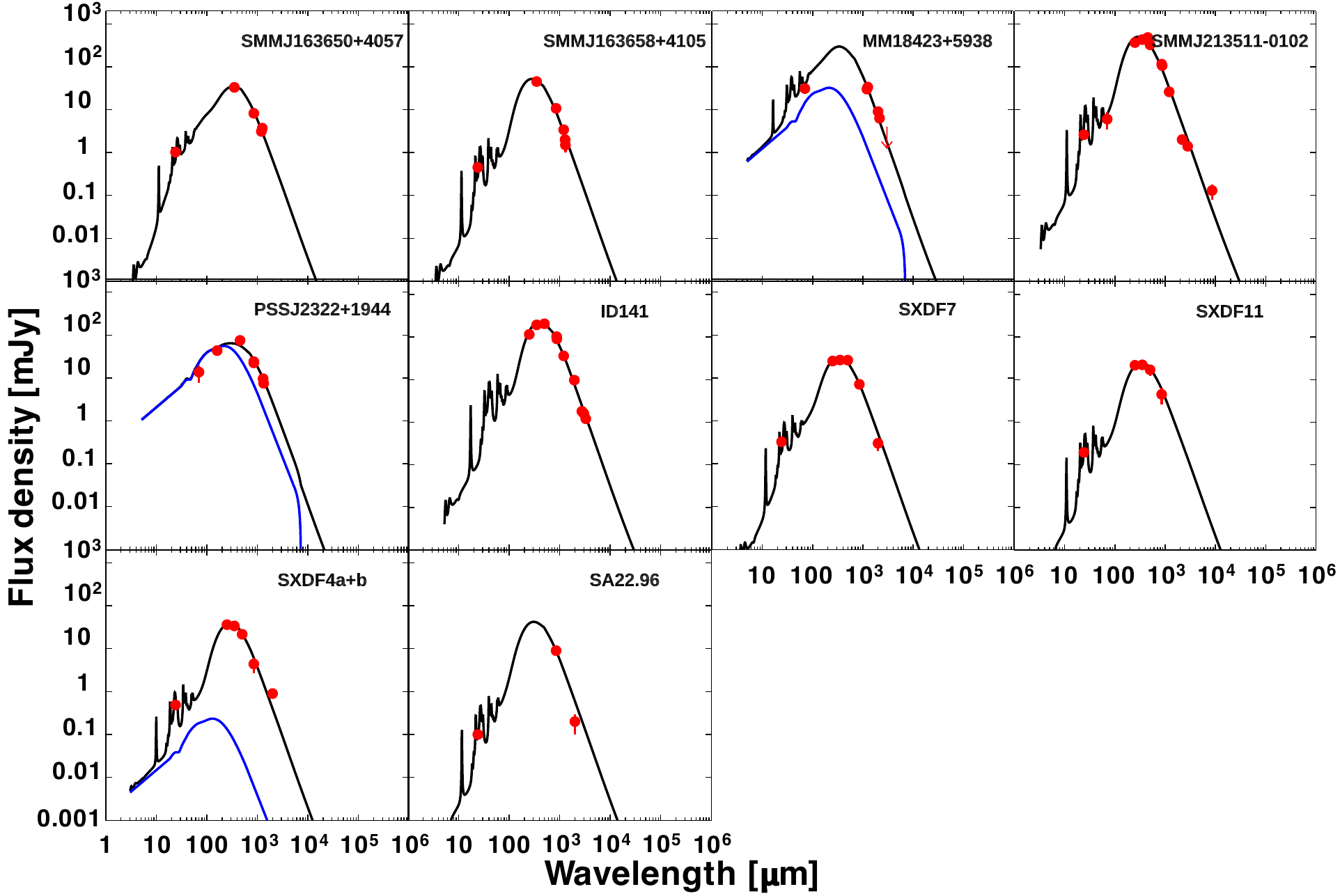}
  \caption{(continue)}
\end{figure*}
\setcounter{figure}{6}    
\begin{deluxetable}{ccccc}
  \tabletypesize{\normalsize}
  \tablecolumns{3}
  \tablecaption{Flux corrections for the ALMA
    observations of galaxies on the main sequence.\label{tab:app:totflux}}
  \smallskip
  \tablehead{
     \colhead{ID}& 
     \colhead{Size}& 
     \multicolumn{3}{c}{Flux increase}\smallskip\\
     \colhead{}& 
     \colhead{\small arcsec}& 
     \parbox{2.5cm}{\centering \footnotesize \cione, \\ \cofour}&
      \colhead{\footnotesize \cofive}&
       \colhead{\footnotesize \cotwo}\\
       \colhead{(1)}&
       \colhead{(2)}&
       \colhead{(3)}&
       \colhead{(4)}&
       \colhead{(5)}
     }
   \startdata
           4233& $1.229\pm0.279$& --& --& --\\
          7540& $1.219\pm0.322$& --& --& --\\
         13205& $<0.448$& $<5$\%& $5$\%& $9$\%\\
         13250& $0.907\pm0.476$& --& --& --\\
         18538& $0.514\pm0.062$& --& --& --\\
         18911& $<1.248$& $31$\%& $113$\%& $39$\%\\
         19021& $0.184\pm0.055$& --& --& --\\
         26925& $1.239\pm0.206$& --& --& --\\
         30694& $0.691\pm0.155$& --& --& --\\
         32394& $1.572\pm0.204$& --& --& --\\
         35349& $0.859\pm0.029$& --& --& --\\
         36053& $<0.578$& $<5$\%& $28$\%& --\\
         36945& $<0.595$& $<5$\%& $<5$\%& $11$\%\\
         37250& $0.893\pm0.044$& --& --& --\\
         37508& $<0.208$& $<5$\%& $<5$\%& $<5$\%\\
         38053& $0.612\pm0.239$& --& --& --\\
         44641& $0.604\pm0.190$& --& --& --\\
        121546& $<1.074$& $19$\%& --& --\\
        188090& $0.426\pm0.015$& --& --& --\\
        192337& $0.579\pm0.026$& --& --& --\\
        208273& $0.459\pm0.048$& --& --& --\\
        218445& $1.152\pm0.347$& --& --& --\\
        256703& $0.891\pm0.030$& --& --& --\\
   \enddata
   \tablecomments{Column 1: ID. Column 2: Size in arcsec. Upper limits
     are at $<1\sigma$. Column 3-5:
     Flux increase when extracting the source with a circular Gaussian
     (\textsc{gildas/uv\textunderscore fit/c\textunderscore gauss})
     with fixed $\mathrm{FWHP}=1\sigma$ upper limit on the size,
     compared with the extraction with a point source profile
     (\textsc{gildas/uv\textunderscore fit/point}): $I_{\rm
       Gauss}/I_{\rm Point}$. Corrections below $<5$\% are not applied.}
 \end{deluxetable}

\section{Total recovered fluxes from ALMA observations}
\label{sec:app:totflux}
 We extracted the flux of each line of our sample
of main sequence galaxies using the iterative procedure described in
detail in our previous work \citep{daddi_2015, coogan_2018,
  valentino_2018, puglisi_2019} and recalled in Section
\ref{sec:ci21_cycle6}. The signal is extracted with
\textsc{gildas/uv\textunderscore fit} at fixed
spatial position and extension, obtained by combining measurements for
each tracer in the \textit{uv} space \citep[Section 2
of][]{puglisi_2019}. Such combination includes
  both cold (\ci, \cotwo, dust continuum) and warm
  (\cofour, \cofive, \coseven) molecular gas proxies. Therefore, the measured
  size is representative of the average extension of the molecular gas
  in the galaxy. Further
comparisons between the various tracers is part of a forthcoming work
(Puglisi et al. in prep.).
The total flux is robustly recovered whenever the
size of the emitting source is securely estimated. This is the case for all
the new Band 7 measurements reported in Section \ref{sec:ci21_cycle6}.
However, flux losses might occur when only an
upper limit on the size can be placed, and such estimate is comparable with the beam
size. We estimated these losses by injecting artificial bright
galaxies with circular Gaussian profiles and a FWHP fixed to the
$1\sigma$ upper limit on the size in the ALMA maps, and then
re-extracting their fluxes with the fiducial point source profile.
We corrected the fluxes of unresolved sources to $(I_{\rm
       Gauss}/I_{\rm Point}+1)/2$ and adding in quadrature the absolute error on such correction ($\sigma_{\rm corr}=(I_{\rm
       Gauss}-I_{\rm Point})/2$) to the statistical uncertainty. The sizes and the flux corrections
are reported in Table \ref{tab:app:totflux}. Changes at $<5$\% are not
significant and therefore not applied. 

\section{Data tables}
\label{sec:app:tables}
The data we collected for this work are released in electronic
\textsc{FITS} format available in the online version of the article or
contacting the corresponding author. We produce two separate tables
for the local and high-redshift subsamples in Section \ref{sec:data}. The quantities are
described in Table \ref{tab:app:columns}.
\begin{deluxetable}{lcl}
   \tabletypesize{\small}
   \tablecolumns{3}
   \tablecaption{Content of the data tables.\label{tab:app:columns}}
   \smallskip
   \tablehead{
     \colhead{Name}& 
     \colhead{Units}& 
     \colhead{Description}
   }
   \rotate
   \startdata
   ID& \nodata& Source identifier\\
   $D$& Mpc& Distance (for the local sample only)\\
   $z_{\rm spec}$& \nodata& Spectroscopic redshift\\
   $\mu$ (dust, gas)& \nodata& Magnification factor for lensed
   sources (for the high-redshift sample only)\\
   \lir\ & \lsun & Total IR luminosity integrated
   within $8-1000$ $\mu$m, corrected for the dusty torus emission if
   the galaxy is not AGN dominated\\
   $T_{\rm dust}$& K& Dust temperature from a optically thin modified black
   body model of the IR emission\\
   $L'$(line)& \kkmspc& $L'$ luminosities of the specified line\\
   $I$(line)& \jykms& Velocity-integrated fluxes of the specified
   line\\
   $f_{\rm AGN}$& \nodata& Fraction of IR emission due to dusty
   tori (for the high-redshift sample only)\\
   Type& \nodata& Galaxy type (MS=Main Sequence; SB=StarBurst;
   AGN/QSO=Active Galactic Nucleus/Quasar; for the high-redshift sample only)\\
   AGN& \nodata& Galaxy with AGN contamination and an entry in
   \cite{veron-cetty_2010} (for the local sample only)\\
    Facility & \nodata&  
   Facility used to detect the \ci\ line emission.\\
    References & \nodata&  
    References to the original works presenting observations of each source.\\
   \enddata
 \end{deluxetable}

\section{Spectra of main sequence galaxies}
\label{sec:app:spectra}
We show in Figure \ref{fig:spectra} the spectra of our sample of $z\sim1.2$ main
sequence galaxies observed with ALMA in Cycles 3 to 6 (Sections \ref{sec:ci10_cycle4}-V18 and
\ref{sec:ci21_cycle6}). The IDs are reported in each panel and
correspond to the ones in Tables \ref{tab:all} and \ref{tab:ci21}. The
black solid lines show the continuum-subtracted spectra in the
observed frame and the red lines mark the best Gaussian fits. The
black ticks signpost the expected line frequency from previous
optical/near-IR redshift estimates. The shaded areas indicate the channels used to
estimate the line fluxes. The number of channels
  covered by each line
  is reported in brackets.
The spectra are color coded as follows: dark green
= \cotwo; blue = \cofour; yellow = \cione, \citwo; orange = \cofive;
purple = \coseven. In case of non-detections, we show the expected
location of the emission with a red solid line, and the channels covered
to estimate the upper limits on the flux as color shaded areas.
 These correspond to the velocity width of other
  significantly detected lines for the same source
  (Section \ref{sec:ci10_cycle4}). 

\section{Tests on PDR modeling}
\label{sec:appendix}

{In Section \ref{sec:modeling} we proposed a simple
interpretation of the observed trends in the line
ratios in terms of PDR modeling. As mentioned, this should be taken as
an order of magnitude estimate, given the limitations of such a simple
approach.
Documented issues arise because of the simple one-dimensional
geometry of classical PDR models, in which the \ci\ emission arises only from a thin layer
within \co\ and \ciplus, while observations suggest full mixing
between \ci\ and \co\ \citep[e.g.,][]{ojha_2001,ikeda_2002}; the
assumption of a micro turbulent medium, not suitable to
reproduce the CO SLED, which is better represented by large velocity
gradients (e.g., LVG modeling, \citealt{young_1991};
\citealt{liu_2015, yang_2017, canameras_2018} for working examples on
subsets of the galaxies in our compilation);
indeed, a clumpy medium, turbulent mixing and cosmic ray (rather FUV-) driven chemistry
radically change this picture and might explain the fully concomitant
\ci\ and \co\ \citep[e.g.][]{papadopoulos_2004, bisbas_2015,
  bisbas_2017, papadopoulos_2018}. More realistic modeling including a
3D geometry and tunable cosmic ray rates resulted in larger gas
densities, but similar radiation fields with respect to the
\citet{kaufman_1999} models adopted here
\citep{bothwell_2017}. The mechanical excitation of CO transitions
through shocks is known to occur in nearby massively star forming
regions and it adds a further layer of complexity not
captured by the simple PDR models we applied
\citep{lee_2019}. We also note that metallicity and carbon abundance
variations among different populations (V18) cannot be captured by the fixed set
of parameters in \citet{kaufman_1999}.\\

We do not apply any more complex modeling to our compilation at this
stage. Notably, we do not make such attempt for our main sequence
galaxies at $z\sim1$ due to the lack of constraining power in our observations. 
The LVG method would likely be able to capture both CO and \ci\
at the same time \citep[e.g.,][]{israel_2015}, but it relies on the
knowledge of both low- and high-$J$ transitions, especially
considering the growing evidence of the
existence of a two-phase ISM, the hottest peaking even above $J_{\rm
  upper}=7$ \citep{liu_2015, yang_2017, canameras_2018}. This
information is not available for the vast majority of our
sample of main sequence galaxies, following the general lack of
constraints on the CO SLED of high-redshift normal galaxies
(\citealt{daddi_2015}, Daddi et al. in preparation).
We refer the reader to the original works producing such LVG modeling when
accessible, mainly for lensed SMGs or very local objects. Similar
considerations apply for the introduction of shock models. Moreover, 
spatially well resolved observations will be necessary to study the
relative distributions of \ci\ and \co\ and gauge, e.g., the effect of
enhanced cosmic rays rates in high redshift galaxies, as unresolved or
marginally resolved emission is dominated by the phases where both
species are abundant \citep{papadopoulos_2018}. We also note 
that the observed ratios \lcitwo/\lci\ $\lesssim3.5$ in all but 3
objects are not compatible with extra heating from X-ray dominated
regions (XDR) at the $n\sim10^4$~cm$^{-3}$ derived for the galaxies
with both \ci\ lines available \citep{meijerink_2007, nesvadba_2018}.

\setcounter{figure}{7}
\begin{figure*}[h!]
  \centering
  \includegraphics[width=0.32\textwidth]{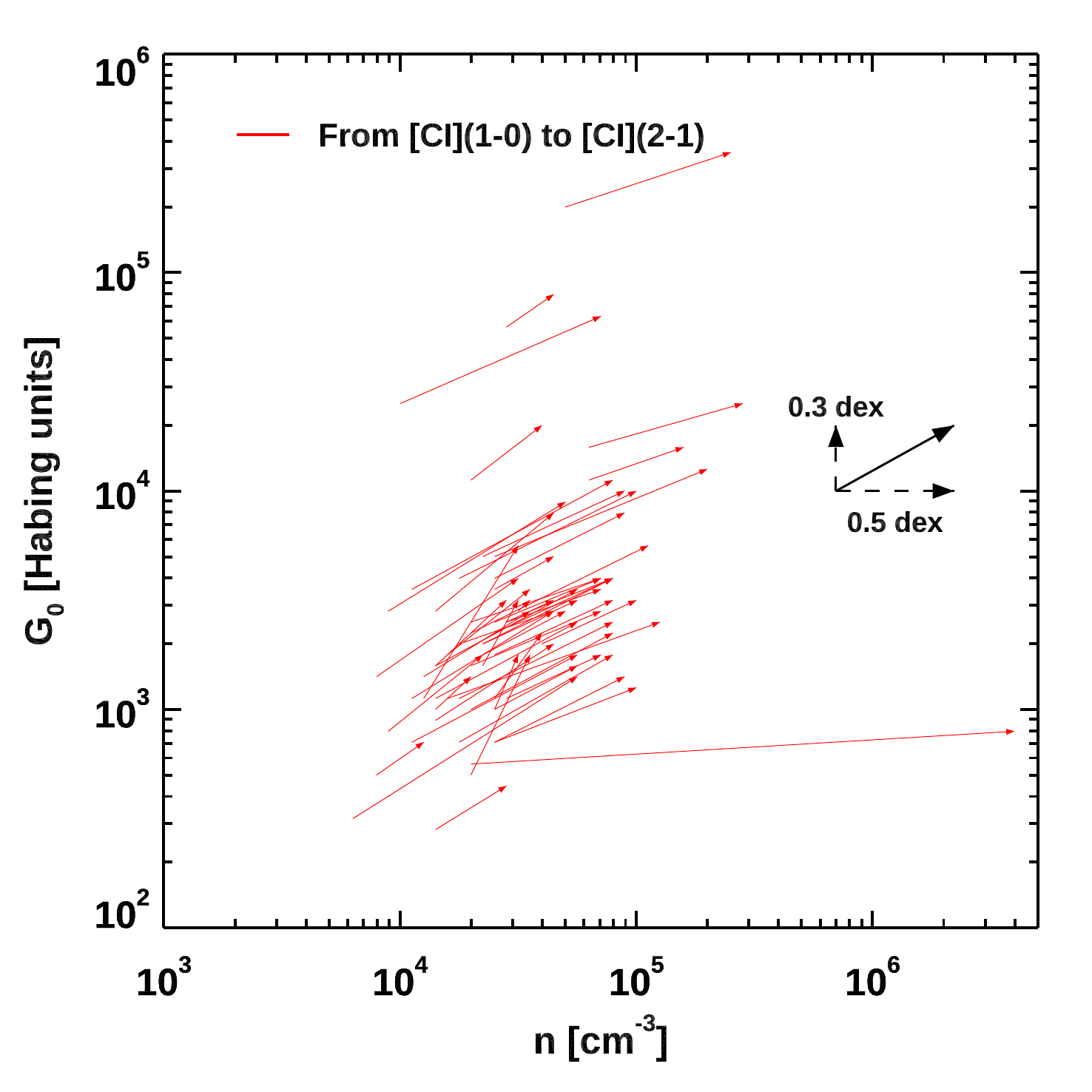}
  \includegraphics[width=0.32\textwidth]{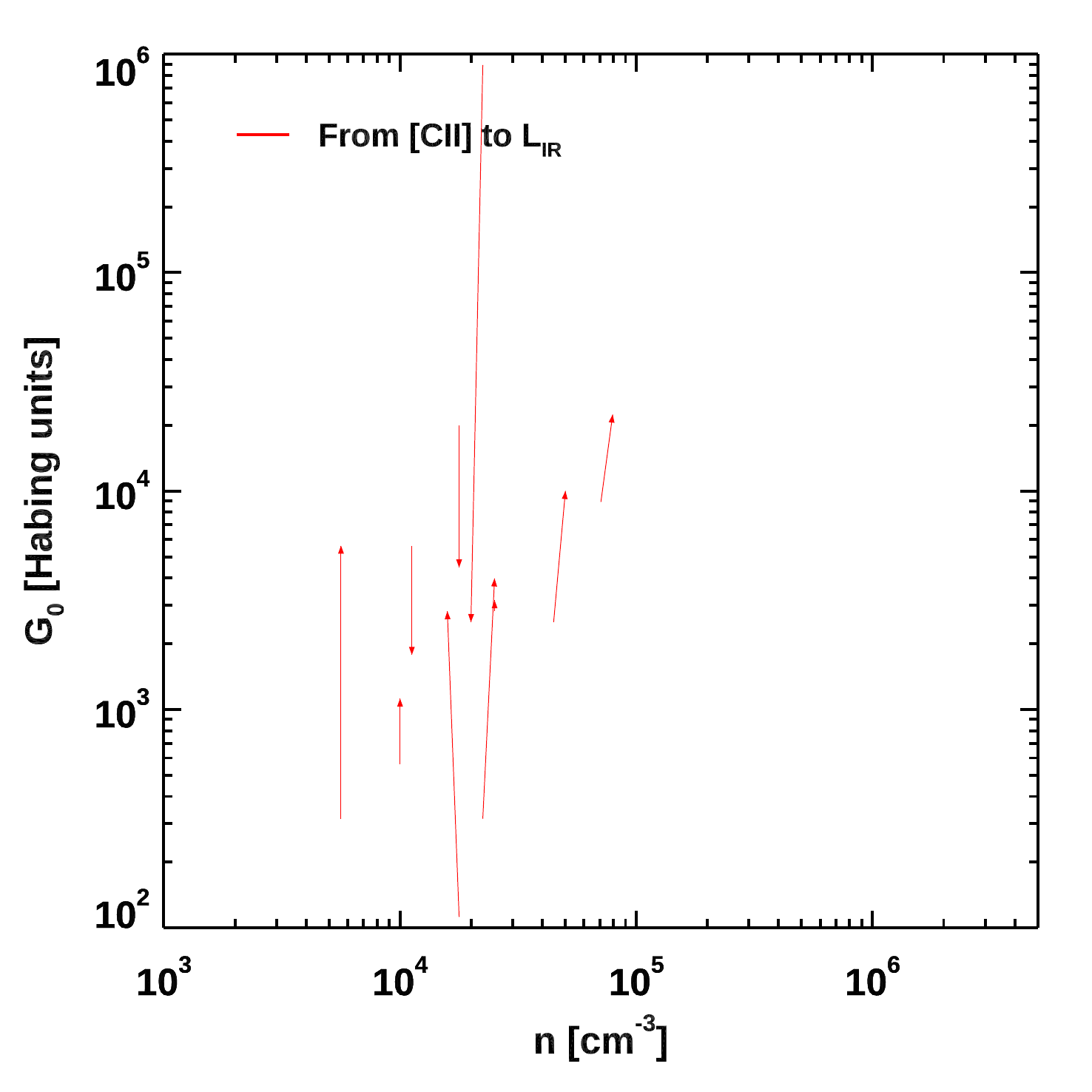}
  \includegraphics[width=0.32\columnwidth]{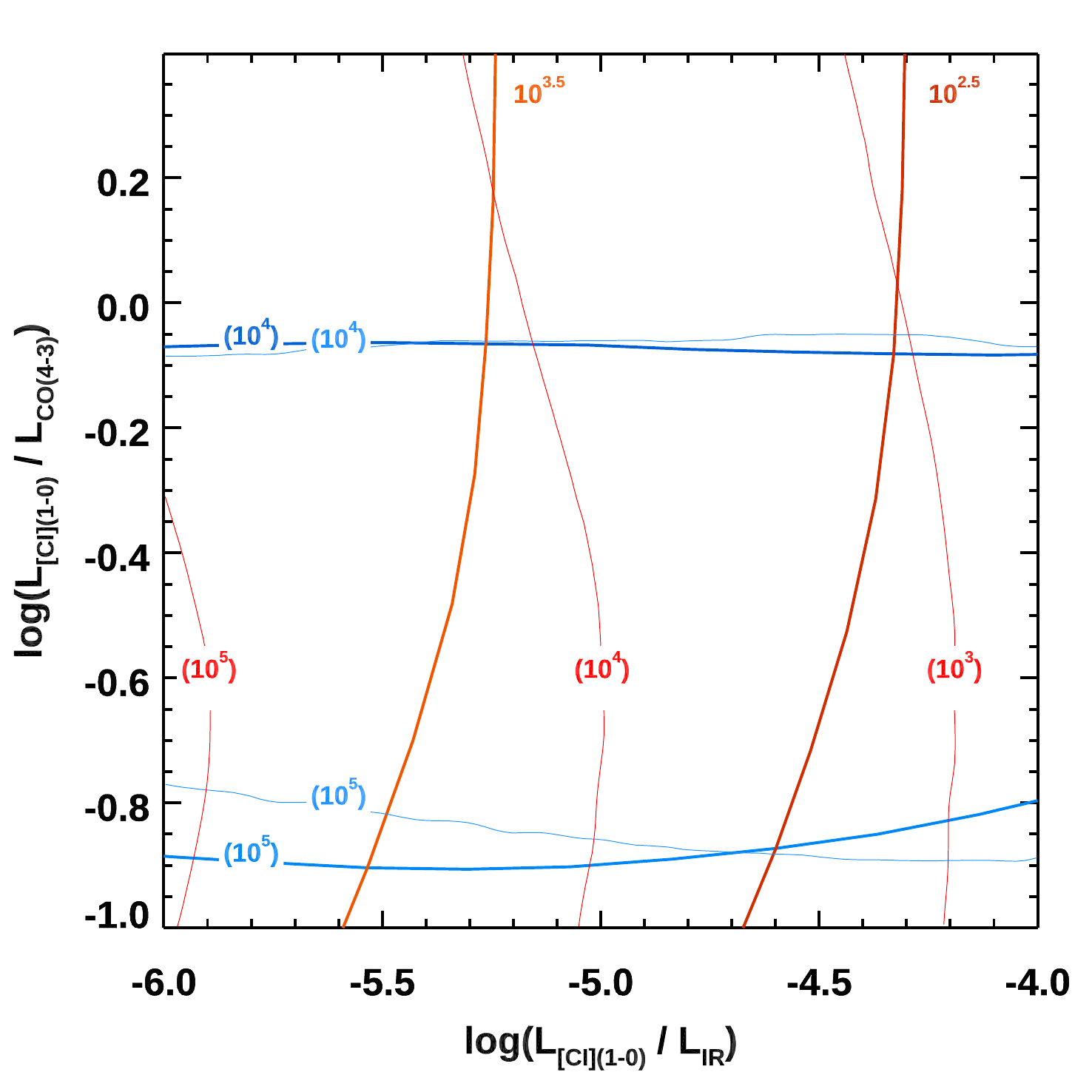}
  \caption{\textbf{Test on \citwo\ vs \cione\ and \ciplus\ and
      comparison with previous modeling.}
    \textit{Left:} Best-fit gas density $n\,[\rm{cm^{-3}}]$ and intensity of
    the UV radiation field $G_{0}\, [\rm{Habing\;  units}]$ from PDR
    modeling by \citet{kaufman_1999} derived using \cione\ or \citwo\ and keeping every other
    quantity fixed. Only galaxies with detections of both lines are
    shown. The red arrows map the difference in $n, G_0$ moving from
    \cione\ to \citwo\ modeling. The black arrows indicate the average difference between
    the two estimates. \textit{Center:} Best-fit $n\,[\rm{cm^{-3}}]$
    and $G_{0}\,[\rm{Habing\;  units}]$ derived using the \ciplus\ or \lir\ as proxies
    of $G_{0}$ and keeping every other quantity fixed. Only galaxies
    with detections of both quantities are
    shown. The red arrows map the difference in $n, G_0$ moving from
    \ciplus\ to \lir\ modeling. \textit{Right:} Blue and red lines respectively indicate
   the tracks of constant gas density $n\,[\rm{cm^{-3}}]$
   and intensity of the UV radiation field $G_{0}\,[\rm{Habing\;
     units}]$ in the \ci, \cofour, and \lir\ plane of Figure \ref{fig:obsratios}.
   The tracks are computed from the PDR modeling by \citet{kaufman_1999}. Thick
   darker lines show the modeling used in this work. Thin lighter
   lines mark the modeling adopted by
   \citet{alaghband-zadeh_2013}. The labels indicate the values
   corresponding to each track.}
 \label{fig:ci_test}
\end{figure*}

\begin{figure*}
  \centering
  \includegraphics[width=\textwidth]{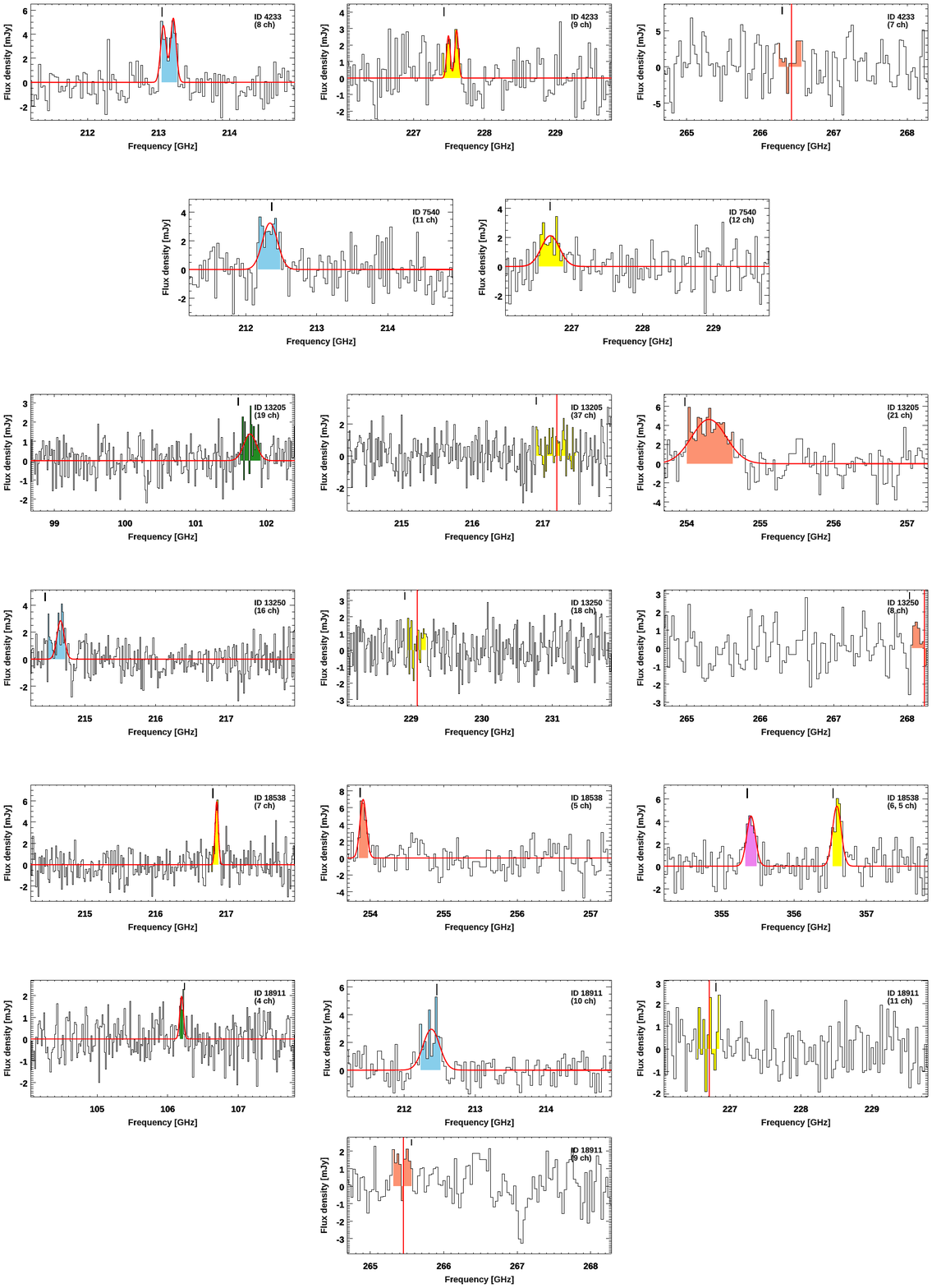}
  \caption{ALMA spectra covering multiple \co\ and \ci\ lines for our
    sample of main sequence galaxies at $z\sim1.2$ 
    (see Appendix \ref{sec:app:spectra} for details).}
    \label{fig:spectra}
\end{figure*}
\setcounter{figure}{8}    
\begin{figure*}
  \centering
  \includegraphics[width=\textwidth]{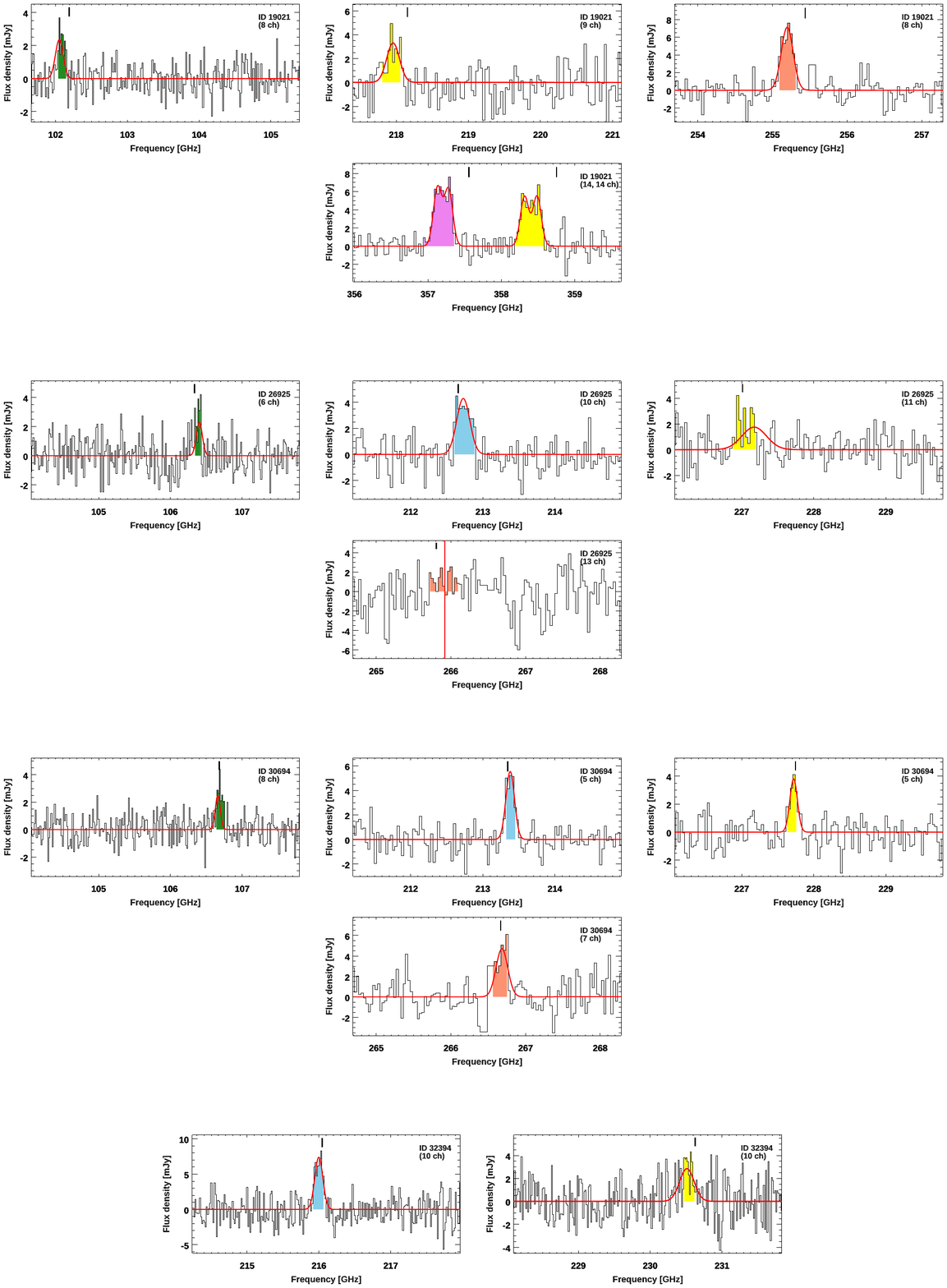}
  \caption{(continue)}
\end{figure*}
\setcounter{figure}{8}    
\begin{figure*}
  \centering
  \includegraphics[width=\textwidth]{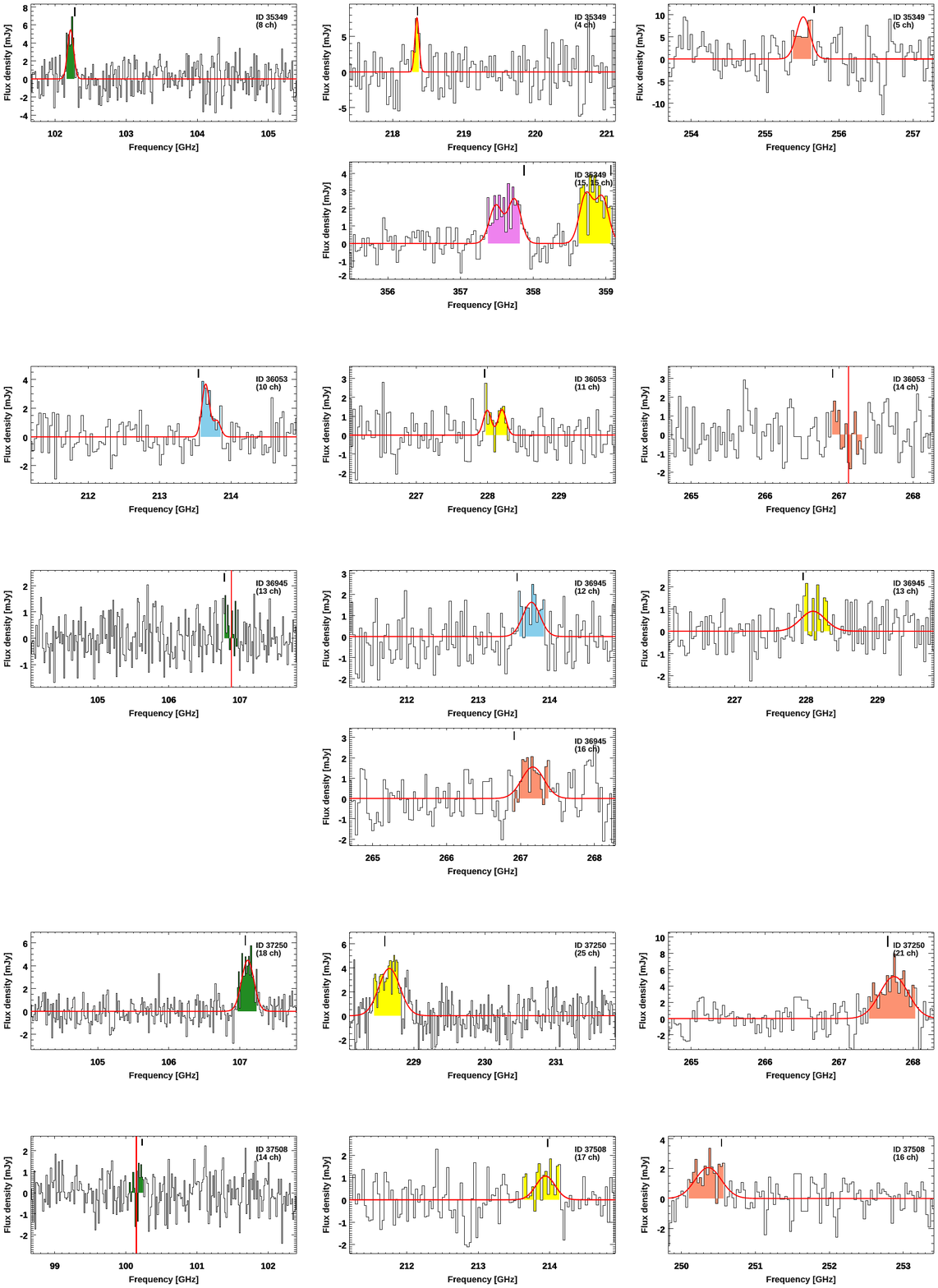}
  \caption{(continue)}
\end{figure*}
\setcounter{figure}{8}    
\begin{figure*}
  \centering
  \includegraphics[width=\textwidth]{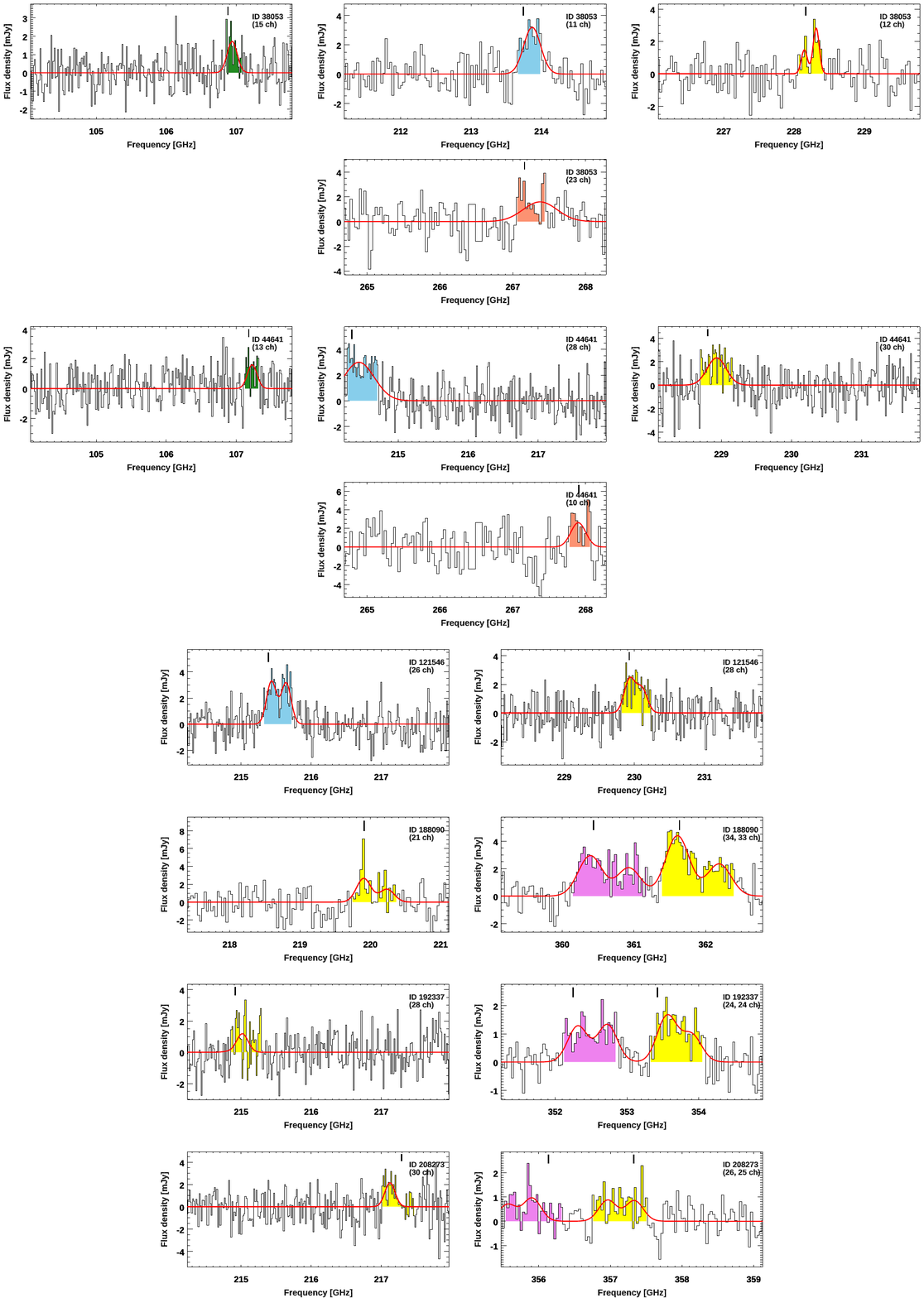}
  \caption{(continue)}
\end{figure*}
\setcounter{figure}{8}    
\begin{figure*}
  \centering
  \includegraphics[width=\textwidth]{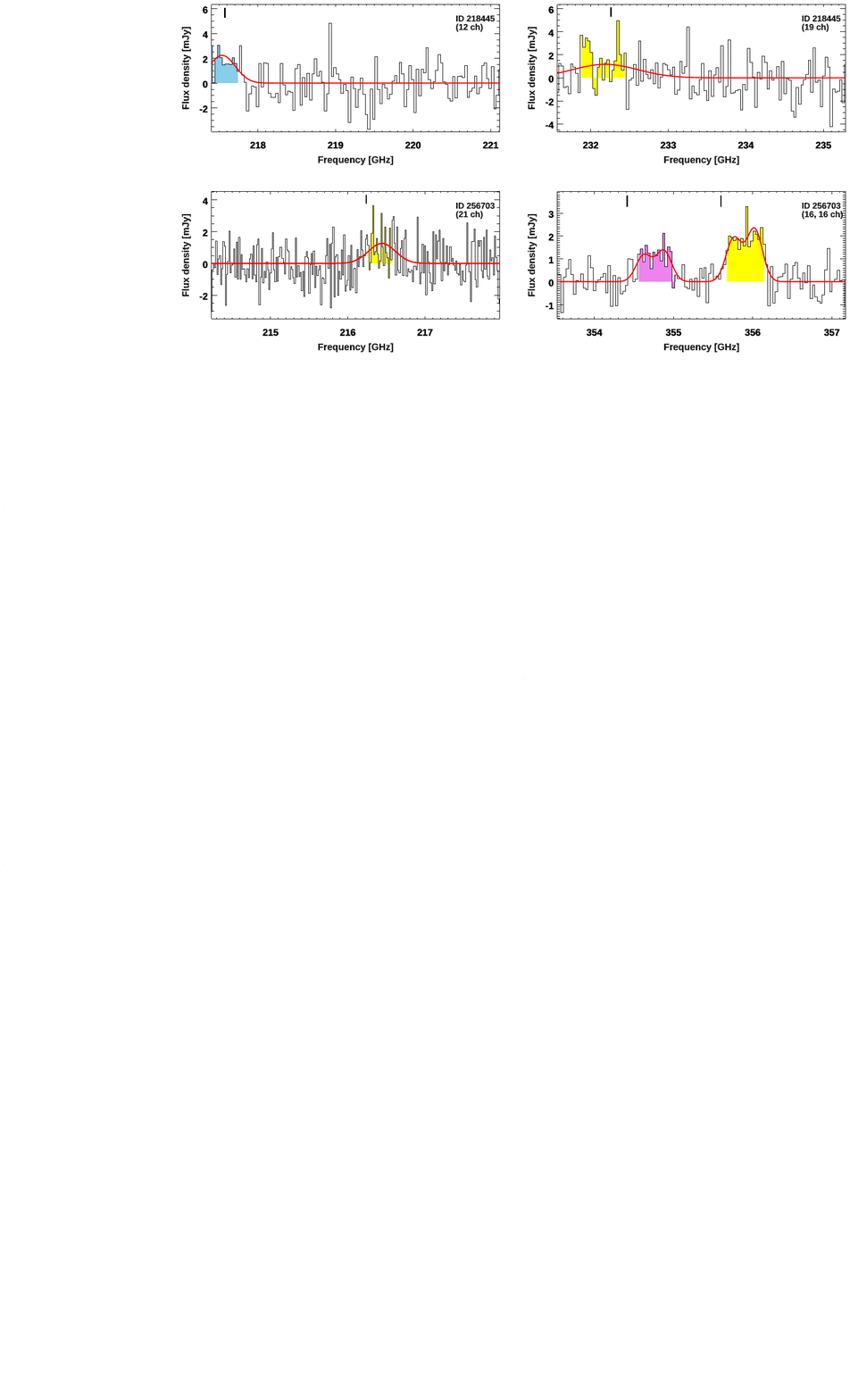}
  \caption{(continue)}
\end{figure*}

\end{document}